# Thick penumbra in a magnetostatic sunspot model

K. Jahn [1,2] and H. U. Schmidt [1]

[1] Max Planck Institut für Astrophysik, Karl Schwarzschild Str. 1, D-85748 Garching, Federal Republic of Germany
[2] Warsaw University Observatory, Al. Ujazdowskie 4, 00-478 Warsaw, Poland

**Abstract.** The conjecture that energy transport in sunspot penumbrae occurs through convection by an interchange of magnetic flux tubes is used in order to construct a magnetostatic model for a deep penumbra which is part of a global sunspot model. The model describes an overall mechanical equilibrium between the three stratifications: of the umbra, the penumbra, and of the solar convection zone, all of them being treated as stellar-like envelopes. Local differences between the gas pressure in the neighbouring stratifications are balanced by the magnetic forces confined to two current sheets: the magnetopause and the peripatopause, whose shapes are determined by the solution of a free boundary problem. A self-consistent coupling between global magnetostatic and thermal equilibrium is achieved by the assumption of a perfect thermal isolation of the umbra and a parametrized inflow of the heat into the penumbra. Characteristic parameters observed in sunspots have been imposed as surface constraints for the model. Then only models with a total magnetic flux larger than about $3 \cdot 10^{21}$ Mx turn out to be feasible. Smaller tubes would require more heat flux than is available in the solar convection zone if they contained an extended penumbra. The solutions obtained indicate that the penumbra contains 50-60% of the total magnetic flux of the spot and has a depth comparable with half the radius of the spot. About 60% of the solar heat flux impinging on the magnetopause seems to be transmitted into the penumbra and conveyed to the photosphere by the interchange convection which can be maintained if the average inclination of the magnetic field exceeds about 30°. Such a penumbra forming an extended buffer causes that the umbra widens with height much less rapidly than the whole flux tube. The efficiency of the convective energy transport in the umbra increases with depth and near the base it can be as strong as in the neighbouring convection zone.

Send offprint requests to: K. Jahn

## 1. Introduction

The traditional picture of a penumbra[1] being a shallow layer formed by an almost horizontal magnetic field seems to be in conflict with observations of the distribution of the magnetic flux in spots. Already the limited radial variation of the field strength across a spot argues against the concept of a flat penumbra. Beckers & Schröter (1969) reduce their measurements of the circular polarized signal with the Unno-Stepanov theory to derive the magnetic field strength. They regard this as the most reliable method to determine the field strength in spots. Their measurements show a moderate decrease of the magnetic intensity across the spot (by a factor of two only) with an abrupt drop to zero outside a penumbra. Also other authors, using different methods of measurements, observe large values of the magnetic field at the outer edge of the spot (e.g. Kawakami 1983, Lites & Skumanich 1990, see also Beckers 1981, and Skumanich 1992). The high field strength at the edge together with force-free expansion of the magnetic field into the Corona leads to the 'magnetic flux problem' (Schmidt 1987, 1991): if the penumbra contained only an almost horizontal field then the total magnetic flux of a spot would have to be equal to the flux through the umbra. This is not the case if one accepts the high value of the magnetic intensity at the outer edge of a spot. A simple estimate of the total and the umbral magnetic flux indicates that the latter is smaller by a factor of about four (Schmidt 1991), if one assumes a decrease of the field strength across the spot by a factor corresponding to the magnetic profile proposed by Beckers & Schröter (1969). This leads to the conclusion that a large fraction of the total magnetic flux should cross the local photosphere in the penumbra. A situation in which the magnetic field becomes horizontal at depths slightly below the visible surface is very unlikely, since there are no outward forces that could balance the sharp local bends of field lines which would have to appear in such a configuration. From the measurements of the magnetic field strength and inclination in spots Solanki &

---
[1] we use the terms 'umbra' and 'penumbra' not only for the photosphere of a sunspot but also for the underlying volume which has a thermodynamic structure different from that of the unperturbed convection zone.

Schmidt (1995) find an observational indication that more than 50% of the total magnetic flux is contained in the penumbra. Therefore the magnetized penumbra over most of its radial extent between the umbra and the quiet Sun should form a thick layer comprising many scale heights. The overall geometry of a spot would then resemble a structure considered by Gokhale & Zwaan (1972), which consists of a central plug, the umbra, surrounded by a mantle, the penumbra, of a width increasing with height (see also Jahn 1989). Such a feature should be taken into account in sunspot modelling, and some attempts to model a thick penumbra are presented in this paper.

Schmidt & Wegmann (1983) presented numerical solutions to the free boundary problem of an axisymmetric magnetostatic flux tube with a given homogeneous pressure stratification separated from the solar convection zone by a sharp discontinuity, the magnetopause. Their method can be used to relate the overall geometry of the magnetic field constituting the flux tube underneath to the magnetic profile at the surface even for a more general internal structure of the axisymmetric tube. A global model for a sunspot, constructed within such an approach, has been utilized in order to probe the stratification of the gas and the configuration of the magnetic field below the photosphere for a given distribution of additional azimuthal volume currents underneath the surface of the penumbra (Jahn 1989). Observational constraints imposed on the model, i.e. the magnetic profile as well as the photometric one, require then that the modelled penumbra contains about 60% of the total magnetic flux and has a thermal structure intermediate between that of the umbra and that of the quiet Sun over many scale heights. The magnetic configuration modelled represents a mean distribution of the magnetic flux such that the overall equilibrium is satisfied, i.e. averaged over inhomogeneities of convective origin in the umbra as well as in the penumbra. This averaging is an inherent feature of such global magnetostatic models of sunspots, and the azimuthal fine structure of the penumbra cannot be described without new substantial extension of this method. The relatively steep inclination of the average magnetic field in the penumbral model is not necessarily in conflict with the Evershed effect which indicates the presence of some almost horizontal field lines. The problem disappears if one accepts that neighbored structures in the penumbra, e.g. bright and dark, contain magnetic fields of substantially different directions, as Beckers & Schröter (1969), or Harvey (1971) proposed. In the global sunspot model discussed here the thick penumbra is approximated by a region permeated by a representative magnetic field of some mean inclination.

Another problem that appears in modelling the structure of sunspots at depth is related to the influence of the magnetic field on the energy transport beneath the surface. That refers not only to a possibly reduced convective efficiency, but also to the general pattern of the heat flow in deeper layers. It may be predominantly vertical, as usually assumed, but probably there are some deviations of the heat flux vector from the vertical direction due to the strong magnetic field which must affect the convective motions, so that one can expect at least a partial channelling of the heat flux, as Hoyle (1949) proposed. This should have an influence on the overall magnetostatic models of sunspots. However, there is no parametrized description of convection in the magnetic field diverging with height we could utilize in the model extending over many scale heights. Therefore, we do not study in detail the nature of the physical processes which control the energy transport inside the spot, but rather apply numerical modelling of the thermodynamic structure of the interior to identify some consequences of the constraints set by photospheric observations and by magnetostatic equilibrium.

Albregtsen & Maltby (1978) reported an interesting correlation of the infrared light intensity ($\lambda 16700$ Å) of umbrae with the phase of the solar cycle, and also the correlation of the mean brightness of the penumbra with the umbral light intensity. In a subsequent series of papers (Albregtsen & Maltby 1981; Albregtsen, Jorås & Maltby 1984; Maltby et al. 1986) these relations were confirmed and studied in more detail. The results indicate a lack of correlation of the spot's brightness with the maximum field strength, the age of spots, their lifetime, size, type, nor with the hemisphere on which spots appear. An extension of these measurements to a wider range of wavelengths (Albregtsen & Maltby 1981) leads to the conclusion that these differences of intensity of umbrae are caused by the variation of the umbral temperature during the solar cycle. Detailed semi-empirical models of the umbral atmosphere (Maltby et al. 1986) show that the temperature of umbrae in the latest phase of the solar cycle is by 300 K higher than the temperature of early phase umbrae. It is also concluded that the mechanism responsible for this temperature variation acts on the whole spot (Albregtsen, Jorås & Maltby 1984), since the mean brightness of penumbrae correlates very well with the umbral brightness.

Schüssler (1980) adapted a working hypothesis that the variation in sunspot brightness is caused by a difference in age of the basic flux tubes forming activity centers. We follow this concept and apply its consequences in modelling the overall structure of a spot. A slow increase of sunspot brightness with the phase of the solar cycle may indicate a relatively slow and inefficient heating of the interior of a flux tube by its slightly warmer environment. This may be supported by the detection of a small increase of the light intensity in a long living spot during its second passage on the solar disk reported by Albregtsen & Maltby (1981). Thus, we may assume that at least the innermost parts of flux tubes are thermally well isolated from their surroundings during the phase of their lifetime, in which they cross the photosphere and form sunspots. This is then the time scale relevant for our magnetostatic models. Consequently, the luminosity of a spot (defined as the heat flux integrated over the cross-section of the tube) should be essentially the same at each depth. This, in turn, is related to the concept which explains the coolness of sunspots by a channelling of the heat flux along the magnetic field lines (Hoyle 1949). Global models for sunspots based on this concept were studied by Chitre (1963) and Chitre & Shaviv (1967). They argued that the inhibition of convection (as proposed by Biermann 1941) as well as the channelling of the heat flux may take place in sunspots. Their model and results are discussed in more detail elsewhere (Jahn 1992).

In the present study of a magnetostatic model of a sunspot we allow for a divergence of the heat flux with height consistent with a heat flow in 1-D stratifications of matter surrounded by current sheets that are thermal insulators. The inner current sheet surrounding the umbra is assumed to insulate perfectly, whereas the magnetopause has a parametrized transmittance and allows for some heat inflow into the penumbra from the convection zone. Thus, the current sheets determine the divergence of the heat flux and also balance the horizontal differences between gas pressure in the neighbouring stratifications, so that the thermal and the magnetostatic equilibrium are self-consistently coupled with each other in such a tripartite model.

In the next section we discuss a hypothetical interchange convection and confront its anticipated signatures with the observed features. Section 3 describes the construction of the tripartite magnetostatic model of a sunspot. In Sect. 4 we discuss the results of the investigation of the proposed model and its stability. Section 5 identifies some conclusions.

## 2. Interchange in a thick penumbra

Radially elongated filamentary structure of penumbrae, and the low radial variation of their mean brightness were explained by field-aligned horizontal convection rolls (Danielson 1961). This seems acceptable for a shallow penumbra but not for a thick one. Measurements of the magnetic intensity and the field inclination in spots favor rather a thick penumbra. Some of the recent findings are discussed below.

Adam (1990) presented observations of the magnetic field vector in a spot which show that even the limiting inclination at the outer edge of the spot amounts to about $20°$ to the horizontal. Her measurements of the inclination angle are based on a method independent of measurements of the light intensity ratios between magnetically split lines. Some average inclination of the magnetic field is determined by means of an analysis of shapes and orientations of polarization ellipses of Zeeman $\sigma$ components only. Therefore, although the field strength in the outer parts of a spot cannot be determined with confidence (due to the overlapping of Zeeman components in a weaker field) the measurements of the field orientation can cover the whole spot. Those measurements show a linear dependence of the tangent of the average inclination angle on the radial distance with a limiting value of $20°$ to the horizontal at the outer edge of the penumbra (see Fig. 5, Adam 1990). Also earlier observations, made with the same but less refined method (Adam 1975), indicate that the average field inclination in the penumbra is not horizontal but varies from $40$–$50°$ at the umbra–penumbra boundary to $20$–$30°$ in the outermost parts of a spot.

Lites & Skumanich (1990) observed the magnetic field in sunspots with the Stokes II polarimeter. Analysis of Stokes profiles of an Fe I line was used to determine the field configuration in four large spots. The authors conclude that the magnetic intensity at the outer edge of the penumbra does not fall below $700$–$800$ G, which is about one third of the central field strength. The average inclination of the field lines is equal to $45°$ at the umbra-penumbra boundary, and has a value of $20°$ at the outer boundary of the spot. The results essentially concur with the results of Kawakami (1983), who analyzed photographically recorded spectra of full Stokes profiles and derived somewhat lower values of field strengths and inclination angles.

Low resolution observations like those cited above give an idea about the overall configuration of the spot magnetic field spatially averaged over relatively large areas. The range of field strengths and inclinations derived favors a picture of a spot with a thick penumbra on phenomenological grounds. Questions concerning the physical nature of the penumbra and the reason of its filamentary appearance require studies of its fine structure.

Lites et al. (1990) performed Stokes profile analysis at high spatial resolution. They found fluctuations of the field strength in the penumbra which can amount to 400 G, and some indication of a rapid change of field inclination between bright and dark filaments. From the spatial variations of Stokes-V spectra Degenhardt & Wiehr (1991) estimated the fluctuations of the inclination angle in the range from $7°$ to $13.5°$, and also found an indication for upward motions at the location of steeper field, which concurs with the suggestion of Beckers (1981) that bright matter moves upwards with respect to dark structures. In neither of those studies a clear correlation of the continuum intensity with the field inclination nor with the Evershed effect was detected. Such correlations have been reported by Title (1989). However, Wiehr & Degenhardt (1992) note that a correlation of the Doppler shift due to the Evershed flow with the light intensity occurs in those of their spectra which achieve exceptionally high spatial resolution. It is very likely that a similar correlation for the magnetic signal might also be present.

A comprehensive study of the fine structure of the magnetic and the velocity field in the penumbra has been presented by Title et al. (1992, 1993) who measured the line of sight magnetic field strength, the intensity, and the velocity signal in a very round spot. These observations show that the average inclination of the field lines varies radially from $45$–$50°$ to $15$–$20°$ across the penumbra. On small azimuthal distances the inclination angle shows fluctuations by about $15°$. Those fluctuations preserve a radial coherence, i.e. they are consistent with the existence of elongated filaments of homogeneous inclination. They also found a correlation of the inclination with the light intensity: the inclination angle is larger in bright and smaller in dark filaments. The Evershed velocity largely occurs in dark penumbral structures so that it is restricted to regions of nearly horizontal magnetic fields. This mass motion is observed as a systematic outward flow of the dark matter in the outer penumbra. In its inner half an inflow of bright matter into the umbra is visible, including an inward motion of 'bright penumbral grains' observed by Muller (1976, 1981). The existence of such general systematic flows dominating in the inner or outer penumbra respectively has been reported also by Zirin & Wang (1991).

The measurements cited above give quite a detailed picture of the dynamic state at the surface of the penumbra. That allows us to propose a description for the thick penumbra as an extended layer between the umbra and the quiet Sun, in which a nonlocal magnetoconvection operates in the form of an interchange of long and narrow magnetic flux sheets (see Schmidt 1991). Such an interchange of entire flux units between the sunspot's magnetopause and the photosphere would then be responsible for the transport of about seventy percent of the solar heat flux through a thick penumbra, and also for its filamentary appearance. Convective elements in this concept should be identified with elongated magnetic flux tubes or rather sheets, since they would have an azimuthal thickness much smaller than their extent in height. The flux elements gain buoyancy as they are heated at the magnetopause. Hotter and less dense matter moves upward towards the photosphere and the umbra. The flux unit emerging through the surface preserves a radial coherence and forms a bright filament in which the magnetic field is significantly inclined to the horizontal. The process of emergence as observed in white light manifests itself as a general inflow of bright matter into the umbra (Zirin & Wang 1991; Title et al. 1992). That includes the inward motion of bright penumbral grains (Muller 1976, 1981) formed most probably by the 'heads' of magnetic units approaching the umbra. In spectral lines the Doppler shift indicating the upward velocity of bright matter is detected (Beckers 1981; Degenhardt & Wiehr 1991).

Emerging flux sheets replace the overlying older elements which had irradiated energy and contain cooler and denser gas so that they sink or are expelled towards the magnetopause. Such a sheet will loose its buoyancy first at its outermost part. Since the hot elements bring to the surface not only the energy but also the magnetic flux (with a more vertical field) the cool ones should remove it. Thus, as the new element proceeds upwards and inwards the old magnetic field becomes more horizontal. Consequently, a subset of almost horizontal field lines is formed, which may satisfy all conditions necessary for the siphon flow (Meyer & Schmidt 1968) so that the Evershed effect is observed.

The process of interchanging the flux sheets is strongly nonlocal and should involve the magnetized plasma extended over many scale heights. They must avoid somehow crossing each other – otherwise convection of this type would be stopped. The Lorentz forces tend to minimize the curvature of the field and smooth out any sharp bends in the field lines. It is conceivable therefore that very long magnetic flux sheets participate in the interchange motion, and that this motion preserves a coherence down to large depths so that long sections of the convective elements are either in the phase 'inward/upward' or in that of 'outward/downward' movement.

This interchange can be maintained only if the buoyancy of the heated elements provides a transverse force large enough to overcome magnetic forces and to pull the field lines inwards. That depends on the average inclination of the magnetopause which, in turn, is related to the total magnetic flux of the tube: the larger the total flux of the spot the less vertical is the magnetopause. Thus, one may argue that for a certain critical magnetic flux the inclination of the field lines at the magnetopause exceeds a critical value at which this form of convection may set in, so that a sunspot with a penumbra, and not a pore, is formed.

Of course, the flux tubes emerging in complex sunspot groups can be strongly perturbed by neighbouring flux concentrations. The magnetopause may then have substantially different inclinations on different sides so that parts of penumbrae of large spots may be missing, or rudimentary penumbrae may form adjacent to moderate flux concentrations. Observations of sunspot formation (e.g. Bumba & Suda 1984) clearly indicate that penumbrae do not develop in the areas where the umbra is directly adjacent to magnetic structures of the same magnetic polarity, that is in regions where we expect the magnetopause to be more vertical.

The inclination of the magnetic field also affects the inward penetration of the interchanging flux sheets. A hot flux unit becomes more and more vertical as it moves inwards, and eventually the transverse component of the buoyancy driving this inflow becomes too small to compete with the magnetic forces. Then the lateral motion ceases rather abruptly as indicated by the sharp transition from the penumbra to the umbra.

The penumbra irradiates about seventy percent of the normal solar heat flux. This energy must be transmitted from the unperturbed convection zone underneath into the magnetized penumbra at a rate sufficiently high, which would be hard to achieve across a plane sharp magnetopause by radiative diffusion. The transmissivity of the magnetopause is directly related to its structure, a problem which we do not address here in detail. We can only speculate that a kind of a diffused magnetopause might be formed, e.g. as a result of the nonlinear development of the fluting instability. The fastest growing modes are those of the shortest azimuthal wavelengths, according to the estimate of Bernstein et al. (1958), and perhaps such a fine-scale interchange might increase the transmissivity of the magnetopause, so that it conveys an essential part of the impinging vertical heat flux.

Our investigation of the magnetostatic model for a sunspot is related to the stable phase of a lifetime of a spot with a regular penumbra that already found a state of an overall equilibrium between the umbral matter and the solar convection zone. We assume that the thermodynamic structure of the penumbra can be approximated then by a 1-D stratification, and describe the transmissivity of the magnetopause with a simple parametrization.

## 3. Construction of the model

A large single sunspot of a regular shape and a long lifetime can be approximated by a unipolar, axially symmetric magnetic flux tube being in mechanical equilibrium with the surrounding convection zone. Applicability of magnetostatic representation of spot's overall structure has been discussed in detail elsewhere (Jahn 1992). The model considered here is based on the approach proposed by Schmidt & Wegmann (1983), who solve a free surface problem for the magnetopause surrounding a force-free magnetic field of the tube. They relate its shape to the difference between the gas pressure outside and inside the tube. The specific model presented here is complementary to the previous model, in which a thick penumbra is taken into account as a region inside the flux tube permeated by electric volume currents (Jahn 1989). Instead we assume here the existence of an additional free surface in pressure equilibrium between the umbra and penumbra. This surface, introduced for mathematical convenience, together with the assumption of a force-free magnetic field, makes it possible to construct a simplified but fully self-consistent approximation for the global sunspot's structure, which can be described by a set of parameters rather than by trial functions.

In the present description of a penumbra we use some of the anticipated features of the interchange convection. First, the lateral mixing of gas by inward and outward moving flux sheets should efficiently smooth out the horizontal gradients of thermodynamic parameters, when averaged over time and azimuth. This is observed at the surface: the mean brightness of penumbrae exhibits a very small lateral variation of about 5% only between the outer and inner edge. It is conceivable therefore, that the lateral gradients of the temperature and density inside a deep penumbra are small, so that this part of a flux tube can be approximated by a force-free magnetic field. Therefore, the continuous distribution of electric currents, which in the previous model contributed to the lateral balance of forces (Jahn 1989), is replaced by a second infinitesimally thin current sheet. This surface, separating the penumbra from the umbra, has been named a peripatopause for it represents a location at which flux sheets peripatetically cease and change the direction of their motion across the penumbra. Thus, the peripatopause approximates the region of the umbra-penumbra boundary, where the filling factor of bright filaments changes rapidly causing the sharp photometric transition in the photosphere. The nonlocal nature of the interchange convection implies that at any depth in a spot roughly the same fraction of the total magnetic flux participates in the motion. The peripatopause should then be aligned with the field lines of the modelled tube, and therefore it is assumed to be a magnetic surface.

The umbral photosphere of regular spots appears to have almost an uniform distribution of brightness, when averaged over small scale inhomogeneities. The variation of the mean brightness is smaller than in the penumbra. We neglect the local relation between the brightness and the field strength observed in spots (e.g. Kopp & Rabin 1992, Martínes Pillet & Vázquez 1993), and assume a lack of any systematic lateral gradients of thermodynamic parameters. This corresponds to the assumption of a current-free configuration of the magnetic field inside the umbra. We use this assumption and approximate the innermost part of the flux tube by a one-dimensional stratification.

As a result of the assumptions made we can represent a sunspot by a simple tripartite model, in which the two magnetic surfaces maintain mechanical equilibrium between the adjacent stratifications that differ from each other. The full model couples self-consistently the overall magnetostatic equilibrium of the flux tube with the thermodynamic structure of the umbra, the penumbra, and of the quiet Sun. The three regions are approximated by one-dimensional stratifications. Each stratification is represented by a plane parallel model of a convective envelope.

Energy transport is described within the framework of the mixing length theory with some additional parametrization, so that a partial inhibition of convection as well as the divergence of the heat flux with height can be taken into account. Differences between the gas pressure in these three regions determine the surface currents which must be taken into account in the solution of the free boundary problem for the magnetopause as well as for the peripatopause.

The model is computed by means of nested iterations. Each of them can be treated as a separate problem, if the necessary input is specified. At the lowest level the thermodynamic structures of the umbra and penumbra are iterated for a given geometry of the current sheets, which determine the channelling effects. The resulting horizontal differences between the gas pressures determine a new distribution of surface currents, which may be inconsistent with the geometry of the current sheets. Therefore, one has to solve again the free surface problem for a new geometry satisfying the magnetostatic equilibrium. One finds a self-consistent solution by repeating iteratively the two steps. We describe the method in consecutive subsections, presenting separately each partial problem solved in the model for a sunspot. An illustration of the configuration and the summary of the full model with all its parameters is given in Fig. 1 and in Table 1. Values given in parenthesis in Table 1 show the range of calculated parameters and refer only to realistic models discussed in the paper.

### 3.1. Free surface problem for two current sheets

The interior of the axisymmetric tube is permeated by a force-free magnetic field and is separated from the field-free plasma by the magnetopause. It is assumed that the magnetic field $\boldsymbol{B}$ may be represented by a scalar potential $u$, so that $\boldsymbol{B} = -\nabla u$. There is a second current sheet inside the tube, the peripatopause. A jump of the magnetic field which tangential on both sheets defines electric surface currents that generate the field throughout the whole space. More details on the mathematical formulation of the present problem, which essentially follows the work of Wegmann (1981, 1987), are given in the Appendix A. Here we briefly outline the method.

The domain considered in the model is defined as a finite section of a flux tube limited by a spherical cup in the upper part and by a plane at the lower end. It is required that the upper and the lower boundary surface normals are everywhere parallel to the magnetic field vector and that the field strength is constant on both limiting surfaces. For reasons of continuity the normal components assume the values of the tangential field at the respective ends of the magnetopause, which have been put far above and deep below the photosphere in order to minimize their influence on the essential part of the flux tube considered.

For the upper cup the constant field strength corresponds to the assumption of a monopole-like configuration, which is well justified by a free expansion of the sunspot's magnetic field into the Corona. The lower end of the flux tube approaches the shape of a vertical column according to the assumption made. Although somewhat arbitrary, this condition is based on the expected tendency of the field lines of quasi-static structures to assume the direction of the gravitational acceleration in the deep convection zone.

**Fig. 1.** Summary of the model parameters shown on the exemplary solution. All parameters are described in the Table 1. The surface of each 1-D stratification is defined by the optical depth $\tau = 2/3$. The variation of the heat flux with depth in the umbra and in the penumbra is represented by means of the shading with an albedo proportional to the value of the heat flux

A jump of the magnetic field at the lower and at the upper boundary defines a distribution of magnetic monopoles which contribute to the potential $u$ generated by the surface currents on the peripatopause and on the magnetopause. The condition that the magnetic field vanishes outside the domain is represented by an integral equation, which states that the potential outside is constant and equal to zero according to the gauging adapted (see also Appendix A). This equation relates the distribution of the sources of the magnetic field to the shape of the boundary, and can be used to determine a position of this free surface. The second requirement, that the peripatopause must be a magnetic surface, implies that the normal component of the magnetic field must be equal to zero there. This condition determines the shape of the second free surface which lies inside the tube. The integral equation is linearized and solved simultaneously with the second condition, if the jumps of the magnetic field at both sheets are given as functions of depth. Then, the overall geometry of the flux tube is determined.

The jump of the magnetic field at the free surfaces is derived from a comparison of gas pressure $P_i(z)$ in the umbra, $P_p(z)$ in the penumbra, and $P_e(z)$ in the solar convection zone. The three stratifications are computed as described in the consecutive subsections. The jump of the field at the magnetopause is given by:

$$B_t(z) = \sqrt{8\pi(P_e(z) - P_p(z))}, \tag{1}$$

and at the peripatopause by:

$$\Delta B_t(z) = \sqrt{8\pi(P_p(z) - P_i(z))}. \tag{2}$$

These two functions define the distribution of the currents on both current sheets and the density of magnetic monopoles on the upper and lower boundaries of the domain. With $B_t(z)$ and $\Delta B_t(z)$ given we solve the free surface problem as described in the Appendix A, and obtain the geometry of the current sheets, which is parametrized by the total magnetic flux $\Phi$, the relative flux that constitutes the penumbra $\phi_{\text{pen}}$ being a fraction of the total flux, and also by the position of the lower ($z_b$) and of the upper ($z_t$) end of the magnetopause. The total magnetic flux $\Phi$ and the lower end of the magnetopause or the bottom of the spot ($z_b$) are the free parameters of the full model. The fractional penumbral flux $\phi_{\text{pen}}$ has to be adjusted in such a way that the radius of the umbra is half of the radius of the spot as it is observed. The exact position of the upper end of the magnetopause, although it is a free parameter, turns out not to be very influential when placed higher than about 500 km above the photosphere. We have chosen $z_t = 800$ km for all our models.

**Table 1.** Summary of model parameters

| Symbol | parameter | type | values | comment |
|---|---|---|---|---|
| $\Phi$ | magnetic flux – total | ○ | $10^{21}, 5 \cdot 10^{21}, 10^{22}, 2 \cdot 10^{22}$ Mx | — |
| $\phi_{\text{pen}}$ | magnetic flux – penumbra | c | $(0.47 \div 0.56)$ | dimensionless fraction of the total flux $\Phi$ |
| $R_p$ | radius of the penumbra | c | $(6.6 \div 20)$ Mm | at the photosphere of the penumbra |
| $R_u$ | radius of the umbra | □ | $0.5 R_p$ | at the photosphere of the umbra |
| $W_p$ | Wilson depression-penumbra | c | $(100 \div 160)$ km | — |
| $W_u$ | Wilson depression-umbra | c | $(460 \div 515)$ km | — |
| $z_t$ | top of the model | □ | 0.8 Mm | upper end of the magnetopause |
| $z_b$ | bottom of the model | ○ | $-23, -20, -17, -15, -13, -12$ Mm | — |
| $z_{bp}$ | bottom of the penumbra | c | $(-2.5 \div -6.2)$ Mm | so that the stratification of the penumbra fits the umbra |
| $\alpha$ | mix. length/press. scale-height: | | | |
| $\alpha_\odot$ | of the quiet Sun | □ | 1 | — |
| $\alpha_{tu}$ | of the upper umbra | c | $(0.15 \div 0.16)$ | above $z_\alpha$ ($z_\alpha = W_u - 0.5$ Mm) |
| $\alpha_{bu}$ | of the lower umbra | c | $(0.68 \div 0.79)$ | below $z = z_\alpha - H_p$ |
| $\alpha_p$ | of the penumbra | c | $(0.70 \div 0.72)$ | above $z_{bp}$ |
| $F$ | heat flux : | | | |
| $F_\odot$ | of the quite Sun | □ | $6.31 \cdot 10^{10}$ erg cm$^{-2}$s$^{-1}$ | |
| $F_p$ | of the penumbra | □ | $0.75 F_\odot$ | at the Wilson depression, $W_p$ |
| $F_u$ | of the umbra | □ | $0.23 F_\odot$ | at the Wilson depression, $W_u$ |
| $F_{ub}$ | | c | $(0.62 \div 1) F_\odot$ | at the bottom of the model |
| $\epsilon$ | transmissivity of the magnetopause | ○ | 0.6, 0.65, 0.7, 0.75, 0.8, 0.9, 1; $\epsilon = 0$ | above $z_{bp}$; below $z_{bp}$ |
| $B_c$ | magnetic field strength-surface | □ | 3000 gauss | at the centre of the umbra ($W_u$) |
| $B_b$ | magnetic field strength-bottom | c | $(9 \div 16)$ kilogauss | at the bottom of the model ($z_b$) |

parameter type: ○ = free, □ = fixed, c = calculated

The solution of the free boundary problem determines the overall configuration of the magnetic field, so that one can calculate some of the parameters characterizing the full spot model once the Wilson depression is known: the central field strength, the magnetic profile at the surface of the tube, the radius of the umbra $R_u$, and that of the whole spot, $R_p$. Moreover, the shapes of the peripatopause and of the magnetopause determine the heat flux divergence, which is used in computing the stratifications of the gas in the umbra and in the penumbra.

### 3.2. The umbra

The envelope that approximates the stratification in the umbra is computed with several auxiliary assumptions concerning the energy transport. They allow for a partial inhibition of convection with an efficiency varying with depth, and also to parametrize the effects of channelling of the heat flux in a simple way.

Radiation alone cannot supply the whole energy irradiated from the umbra, so that there is some form of convective transport just below the visible surface. This fact is indicated by empirical models of the umbral atmosphere, which appear inconsistent if a purely radiative gradient of the temperature is assumed for $\tau \geq 1.5$ (Zwaan 1968, 1974, 1975; Van Ballegooijen 1981; Maltby 1992). Convection is also required in global sunspot models (e.g. Deinzer 1965, Jahn 1989), since otherwise the umbral heat flux would be about ten times smaller.

There is no rigorous theory of heat transport by magneto-convection that could be applied quantitatively in our global model which extends over many scale heights. Therefore, the temperature gradient in the umbra is determined by means of the mixing length formalism with additional assumptions which follow the results of the previous study of the subphotospheric structure of a sunspot (Jahn 1989). His model, which fits the observed magnetic and the photometric profile at the surface, requires a stratification of the umbra which is characterized by a particular gradient of the temperature. This gradient was used to calculate the mixing length parameter $\alpha = l/H_\mathrm{p}$ ($l$ being the mixing length, and $H_\mathrm{p}$ the pressure scale height) at each depth. The resulting variation of the mixing length parameter with depth exhibits indeed an abrupt increase by a factor of 2–3 at depths of about 1500 km below the surface. That concurs with the local analysis of convective stability applied to a sunspot model (Meyer et al. 1974): the efficiency of magnetoconvection is larger in deeper layers, due to the variations of the magnetic ($\eta$) and thermal ($\kappa$) diffusivities with depth. Down to about 2000 km below the solar surface, where $\eta < \kappa$ in their model, only less efficient oscillatory solutions are found. Below that depth the magnetic diffusivity is larger than the thermal one $\eta > \kappa$, so that the plasma can leak across the magnetic field and some kind of overturning convection can develop.

Weiss et al. (1990) studied the effects of gravitational stratification and of variations of the diffusivity ratio $\eta/\kappa$ on compressible nonlinear magnetoconvection. One of the cases investigated corresponds to the situation one expects in the umbra: $\eta/\kappa < 1$ near the top, and $\eta/\kappa > 1$ near the bottom of the domain considered. The results point out the importance of nonlocal effects on magnetoconvection. The two types of instability (oscillatory above and overturning below) are coupled with each other by nonlinear interactions and a time dependent convection develops in the form of hot plumes penetrating to the upper boundary. Such a motion is more efficient than oscillatory convection at transporting energy and probably might account for the relatively large heat flux in the umbra.

We use the hints discussed above and determine the stratification of the umbra in the following way. The uppermost layers at optical depths smaller than 2/3 are represented by a grey atmosphere in Eddington's approximation with a heat flux $F_\mathrm{u}$ constant with depth. The atmosphere of the umbra resembles the atmosphere of the late type stars but it is not the same. In particular, there are no significant convective motions in the upper photosphere of the umbra as they are in a photosphere of a cool star with the same heat flux, where the overturning convection is not suppressed so that the temperature gradient can be smaller. Beckers (1977) gives an upper limit for the convective blue-shift of spectral lines in the umbra equal to 25 m·s$^{-1}$. Therefore, we suppress artificially any convection above $\tau = 2/3$ in the umbra model, whenever the Schwarzschild criterion in our code indicates such instability. Then, the temperature gradient is always radiative in these layers.

The stratification below the surface is convectively unstable due to a large superadiabaticity. Convection is not suppressed in models for realistic stratification in the umbra even if the stability condition modified for the presence of the magnetic field is used (Moreno-Insertis & Spruit 1989). However, there is no parametrized theory of the energy transport in the magnetic field that could be applied to calculate the temperature gradient in those layers. Therefore, we use the results obtained by means of the diagnostic model of Jahn (1989): the gradient below the surface is determined by a solution of the standard mixing equations with a specific variation of the parameter $\alpha$ (see Fig. 1 and Table 1), which mimics the variation determined in the model of Jahn (1989). The parameter $\alpha$ is constant down to a certain depth $z_\alpha$ and has the value $\alpha_\mathrm{tu}$. Below, it increases with depth over the range of less than one scale height where it assumes the value $\alpha_\mathrm{bu}$, and stays constant in still deeper layers:

$$\alpha(z) = \begin{cases} \alpha_\mathrm{tu}, & z > z_\alpha \\ \min\left(\alpha_\mathrm{bu}, \max(|z - z_\alpha|/H_\mathrm{p}(z), \alpha_\mathrm{tu})\right), & z \leq z_\alpha \end{cases} \quad (3)$$

We adapted also the hypothesis that the variation of umbral temperature during the solar cycle reflects the inefficient heating of the flux tubes constituting spots, and assumed that the umbra is thermally isolated from the surrounding. Consequently, all energy irradiated in the umbra model comes from the volume pertained to the umbra.

The effects of channelling of the heat flux has been checked by Deinzer (1965) in his 'similarity-law' models with the conclusion that the funnelling of field lines is too small to account for the low brightness of spots. On the other hand, the models considered by Chitre (1963) and by Chitre & Shaviv (1967) seem to overestimate this effect. The channelling was incorporated in their models by mapping the heat flux vector on the magnetic field vector. This mapping together with the neglect of curvature in the approximate balance of forces overestimates the vertical gradient of the field strength, so that the compressed heat flux reaches the value of the solar flux $F_\odot$ already at depths 2000–3000 km below the photosphere. In neither of the models mentioned above the channelling has been self-consistently coupled with the overall magnetostatic equilibrium.

In the present model the channelling is accounted for by the simple assumption that the luminosity of the umbra, defined by the heat flux $F_\mathrm{u}(z)$ integrated over the horizontal cross-section of the umbra $\pi R_\mathrm{i}^2(z)$, is constant with depth. Consequently, the heat flux varies with depth inversely proportional to the area of the umbral flux tube:

$$F_\mathrm{u}(z) = F_\mathrm{u}\left(R_\mathrm{i}(W_\mathrm{u})/R_\mathrm{i}(z)\right)^2, \quad (4)$$

where $W_\mathrm{u}$ is the Wilson depression of the umbra. We assume the value of the heat flux at the surface $F_\mathrm{u} = 0.23 F_\odot$, which corresponds to the effective temperature of the umbra $T_\mathrm{eff} = 4000$ K. The shape of the peripatopause $R_\mathrm{i}(z)$ is determined by the solution of the free surface problem (see Sect. 3.1.), so that Eq. (4) describes the effect of the global magnetostatic equilibrium of the flux tube on the thermodynamic stratification in the umbra. The heat flow in the model, as implied by our assumptions, is illustrated in Fig. 2.

The upper boundary conditions of the umbral stratification are given as in Eddington's atmosphere: for the gas pressure $P_\mathrm{i}(\tau = 0) = 0$, and for the temperature $T_\mathrm{i}(\tau = 0) = \sqrt[4]{2} T_\mathrm{eff}$. The lower boundary conditions are given at $z = z_\mathrm{b}$, and are specified by the requirement that the spot becomes a vertical column there. The magnetostatic equilibrium requires then that the density inside the flux tube should be equal to that in the convection zone ($\varrho_\mathrm{i} = \varrho_\mathrm{e}$). For the assumed value of the field strength $B_\mathrm{b}$ the gas pressure inside the tube at the bottom $z_\mathrm{b}$ and below is given by:

$$P_\mathrm{i}(z) = P_\mathrm{e}(z) - B_\mathrm{b}^2/8\pi, \quad (5)$$

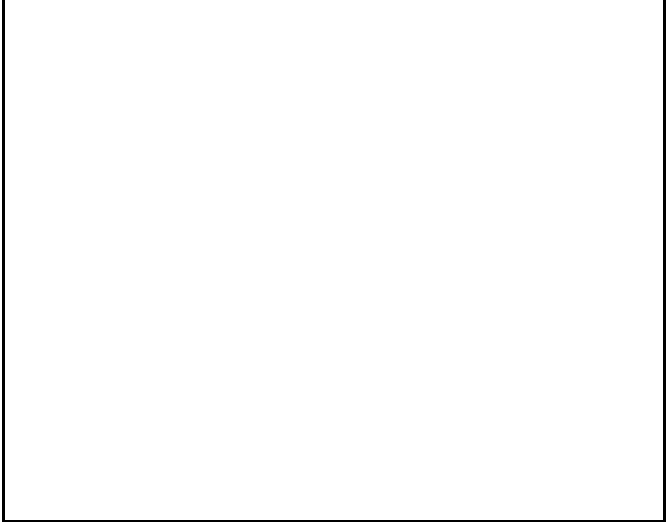

**Fig. 2.** A schematic representation of the assumed heat flow in the sunspot model. The dashed line corresponds to the partially permeable part of the magnetopause. Its remaining part and the whole peripatopause do not transmit energy

where $P_e(z)$ is the gas pressure in the solar convection zone. Since the field strength in a vertical column is constant with depth and the density is the same as outside, the temperature variations with depth in the interior are determined by Eq.(5) together with the equation of state. One finds then, that the logarithmic temperature gradient in the umbra ($\nabla_i = \mathrm{d}\ln T_i/\mathrm{d}\ln P_i$) is related to the external gradient:

$$\nabla_i = (1 - 1/\beta)\nabla_e + 1/\beta, \qquad (6)$$

where $\beta = 8\pi P_e/B_b^2$. The heat flux at the bottom $F_{ub}$ is known (see Eq. (4)). Thus, the full thermodynamic state of the matter can be calculated at $z_b$, including the gradients of thermodynamic parameters. That gives no freedom for the value of the mixing length parameter $\alpha_{bu}$, which is then a function of the following variables: $\alpha_{bu} = \alpha(z_b, F_{ub}, B_b)$.

With both functions $\alpha(z)$ and $F_u(z)$ prescribed (cf. Eqs. (3) and (4); note that the solution for the free boundary must be known) the stratification of the umbra is determined by means of a shooting method. At this step of the computation the values of $z_b$, $z_\alpha$, $B_b$, and in some sense also $R_i(z)$ can be regarded as free parameters, but in the full model of a spot only $z_b$ is a free parameter ($z_\alpha$ is fixed in the whole study). With all these parameters given the values of the Wilson depression ($W_u$) and of the mixing length parameter in the upper umbra ($\alpha_{tu}$) are varied until the downward integration of the stratification yields the density and the gas pressure equal to the boundary values prescribed at $z_b$. Eventually, the thermal structure of the umbra, affected by the overall magnetostatic equilibrium via $R_i(z)$, is determined. In particular, there is derived the gas pressure $P_i(z)$. It is used subsequently in calculations of the magnetic field jump at the peripatopause, i.e. $\Delta B_t(z)$, which is an input function for the free surface problem (cf. Eq. (2)).

### 3.3. The penumbra

The region extending between the magnetopause and the peripatopause is also approximated by a one-dimensional model of a convective envelope. However, the parametrization of the energy transport differs slightly from that for the umbra, since we allow for a thermal contact with the surroundings, so that the luminosity of the penumbra is not necessarily constant with depth.

Interchanging magnetic flux units transport the energy from the magnetopause to the surface of the penumbra. Heating of these flux elements at the magnetopause must be more efficient than a possible radiative diffusion across a smooth interface. Probably a kind of local, small scale convection operates in the magnetopause, but a quantitative description of a relevant mechanism cannot be provided at present. Therefore, we approximate the efficiency of heating by a parameter $\epsilon$ which describes the transmissivity of the magnetopause to the impinging vertical heat flux. Then, the heat flux entering the penumbra through the surface unit of the magnetopause is equal to the solar heat flux reduced by the projection and by the transmissivity coefficient chosen, i.e. it is represented by $\epsilon F_\odot \cos\theta$, where $\theta$ is the inclination of the surface element to the horizontal. Further, in order to allow a one-dimensional description it is assumed that at each depth the whole flux in the penumbra is uniformly redistributed over the full horizontal plane between the magnetopause and the peripatopause. The mean value of the heat flux $F_p(z)$ in the penumbra model depends then on the geometry of both current sheets and on the transmissivity parameter $\epsilon$. Adding both contributions of the heat flux and taking into account the horizontal redistribution one derives at the following variation of the penumbral heat flux with depth:

$$F_p(z) = F_p \frac{R_p^2 - R_u^2}{R^2(z) - R_i^2(z)} - F_\odot \frac{R_p^2 - R^2(z)}{R^2(z) - R_i^2(z)} \epsilon, \qquad (7)$$

where $F_p$ is the mean flux observed in the photosphere of the penumbrae (we take $F_p = 0.75 F_\odot$, which corresponds to $T_{\mathrm{eff}} = 5375$ K), $R_u$ and $R_p$ are the surface radii of the umbra and of the penumbra (see Fig. 1), and $R(z)$ is the radius of the magnetopause.

One can expect that the transmissivity depends on the inclination of the magnetic field and thereby on the depth, but since the relevant mechanism is not known we prescribe simply a constant value of $\epsilon$ above the bottom of the penumbra and do not allow for any heat exchange across the magnetopause below that level. The bottom of the penumbra is defined as a depth $z_{bp}$ at which the stratification of the penumbra merges with the umbra. The heat flow in the penumbra is illustrated in Fig. 2, where the dashed fragment of the magnetopause indicates a partial transmissivity. The value of $z_{bp}$ is calculated by means of Eqs. (4) and (7) with the condition that the heat fluxes $F_u(z)$ and $F_p(z)$ are equal at $z_{bp}$. Thus, the depth of the penumbra depends on the geometry of both current sheets $R(z)$ and $R_i(z)$, and thereby on the overall magnetostatic equilibrium of the flux tube. We assume also that $z_{bp}$ defines implicitly the mini-mum inclination of the magnetopause at which the interchange convection can still be maintained.

The stratification is determined in a similar way as for the umbra. An Eddington grey atmosphere approximates the optically thin regions, with the heat flux $F_p$ and with standard upper boundary conditions (compare Sect. 3.2.). But now the possible convection above $\tau = 2/3$ is not artificially suppressed. The treatment of the mixing length theory and of optically thin convective elements follows that of Paczyński (1969). With the function $F_p(z)$ prescribed the stratification is integrated downwards with a constant mixing length parameter $\alpha_p$, until $z_{bp}$ – the level of merging of the umbra with the penumbra – is reached. There, the lower boundary conditions require that the pressure and the density are the same as in the umbra. We shoot with values of $\alpha_p$ and of the Wilson depression $W_p$ in order to fulfill these conditions.

The thermal structure of the penumbra is governed by the transmissivity parameter $\epsilon$ and by the fraction $\phi_{pen}$ of the total magnetic flux contained between the two current sheets. Of these two only $\epsilon$ is a free parameter of the full model (see Fig. 1 and Table 1). The obtained gas pressure in the penumbra $P_p(z)$, is used to calculate the jump of the field strength $B_t(z)$ on the magnetopause, and $\Delta B_t(z)$ on the peripatopause, cf. Eqs. (1-2). Note, that $\Delta B_t(z) = 0$ for $z < z_{bp}$, because both stratifications become identical there. $\Delta B_t(z)$ vanishes also at heights of few hundred kilometers above the surface of the umbra, where the differences between the gas pressures, which are small compared with the magnetic pressure, become negligible.

### 3.4. The quiet Sun

For completeness we present also the external stratification computed as a standard plane parallel model for the solar convection zone with Eddington's atmosphere of the heat flux $F_\odot = 6.31 \cdot 10^{10}\mathrm{erg\,cm^{-2}s^{-1}}$ ($T_{eff} = 5777$ K), with the mixing length parameter $\alpha_\odot = 1$, and with the solar chemical composition ($X = 0.7331$, $Z = 0.0169$). The latter is applied also to the umbra and penumbra, of course. The equation of state includes effects of partial ionization. We have used Los Alamos opacity tables (Huebner et al. 1977).

Blocking of the heat flux in our sunspot model takes place only at the magnetopause, which is partially permeable in a certain range of depth. It has been assumed that the remainder of the energy impinging on the magnetopause is diverted and distributed over a large volume in the solar convection zone as shown in Spruit's (1977) calculations. Consequently, all possible effects of heat blockage on the stratification adjacent to the magnetopause have been neglected. The standard unperturbed model for the solar convection zone yields the gas pressure variation in the exterior $P_e(z)$ used to derive a jump of the magnetic field at the magnetopause (cf. Eq. (1)).

### 3.5. Iteration of the full model

The configuration of the magnetic field in this model is related to the thermal structure of the umbra and of the penumbra by the local differences between gas pressures which define the distribution of electric surface currents or the jump of the magnetic field on the current sheets. The thermal structure of the flux tube depends, in turn, on the magnetic configuration, since the geometry of both current sheets influences the heat divergence with height. Then, the full model can be obtained with an iterative procedure consisting of two main steps. At first, the three stratifications are computed for the trial geometry assumed, so that the variations of the heat flux can be defined (see Eq. (4) and (7)). Then the jumps of the magnetic field are calculated on both current sheets by means of Eqs. (1)-(2). The electric currents associated with these jumps extend along the whole magnetopause ($j(z) = B_t(z)/4\pi$), while the peripatopause contains such currents only in the region where the stratification of the umbra and of the penumbra are different, i.e. above $z_{bp}$ ($j_i(z) = \Delta B_t(z)/4\pi$). In the second step, the free boundary problem is solved and the geometry of current sheets is determined. Then, the heat flux variations are calculated again and the whole procedure is repeated. The solution is accepted as the final model when the currents and the associated shapes of the free surfaces, no longer change.

There are only three free parameters governing the model: the total magnetic magnetic flux of a spot $\Phi$, the position of spot's bottom $z_b$, and the transmissivity of the magnetopause $\epsilon$. The remaining parameters (see Fig. 1 and Table 1 for a summary) have values either constrained by observations or determined in consecutive iterations. The model is fitted to a 'typical spot', which has the macroscopic properties typical for a large, regular, single spot appearing in the middle phase of the solar cycle. A study of different values of the central field strength, of the relative umbral radius, or of the brightness, though needed, remains beyond the scope of the present paper. The constraints chosen are the following: the central field strength $B_c$ is put equal to 3000 gauss, the radius of the spot is two times larger that the radius of the umbra ($R_u = 0.5R_p$), and values of the heat flux in the umbra and in the penumbra are $F_u = 0.23F_\odot$ and $F_p = 0.75F_\odot$, respectively.

Iterations of the thermal structure of the umbra and of the penumbra yield all values of the mixing length parameters: $\alpha_{tu}$, $\alpha_{bu}$, $\alpha_p$, and of the Wilson depressions $W_u$ and $W_p$. The remaining parameters $z_{bp}$, $F_{ub}$, $B_b$, and $\phi_{pen}$ are determined in the iteration of the full model, which contains several implicit relations. Thus, the field strength $B_b$ at the bottom is related to the value $B_c$ at the surface by the condition of the conservation of the total magnetic flux. The magnetic flux $\phi_{pen}$ that constitutes the penumbra must be such that at the surface the relation $R_u = 0.5R_p$ is fulfilled. Finally, the position of the bottom of the penumbra $z_{bp}$ and the heat flux at the bottom of the model $F_{ub}$ are implied by the coupling of the overall magnetostatic equilibrium with the thermal one.

## 4. Results

The coupling of the global magnetostatic equilibrium with the thermodynamic structure of the flux tube is represented in the tripartite model by a set of parameters rather than by a trial function as in the previous model (Jahn, 1989). For this simplicity, however, the approximation for the surface layers of a spot is rather crude. The treatment of the transitions umbra-penumbra and penumbra-quiet Sun is inaccurate, since all three regions are represented by one-dimensional stratifications, also in optically thin layers. As a result, the models have artificial jumps in all thermodynamic parameters and also in the surface profile of the magnetic field at $r = R_u$, which is not a significant disadvantage of the model as long the global nature of the spot is considered.

Figure 1, used to present all parameters of the model, shows an exemplary solution illustrating typical features of computed models. The horizontal size of the whole flux tube at the surface is at least 1.6 times larger than the size at the base of each model. The associated compression of the magnetic field and of the heat flux at $z_b$, which both scale roughly like an inverse area of the tube, is then about threefold. The channelling effect is shown in Fig. 1 by means of shading with an albedo proportional to the value of the heat flux. The heat flux decreases with height in the whole spot up to the bottom of the penumbra, above which the penumbra is fed with energy from outside so efficiently that its heat flux increases with height despite of the channelling. The umbra, being thermally isolated from the surroundings by the peripatopause, conserves the luminosity and has a flux decreasing with height.

The magnetopause above the photosphere behaves similarly in all models: it continues up to its upper end at $z_t$ which is the fixed parameter (cf. Table 1 and Appendix A). This height is reached at the radial distance of about $3 - 4 R_p$. At the distance $r \approx 2 R_p$ the magnetopause is practically horizontal (with inclinations $4 - 6$ deg) and lies at heights between 400 and 500 km. It is not very accurate, however, to identify the magnetopause in the tripartite model directly with canopies observed in spots, for the reasons mentioned at the beginning of the section.

The overall geometry of the magnetopause and of the peripatopause depends on the distribution of the surface currents. Typical variations with depth of these currents are presented in Fig. 3, by means of the magnetic field jumps $B_t$ and $\Delta B_t$. The dotted line corresponds to the jump $B_t$ at the magnetopause and the dashed one to $\Delta B_t$ at the peripatopause. There is also shown there the total difference between the gas pressure in the exterior and in the umbra, expressed by means of the fictitious field strength $\Delta B_{tot} = \sqrt{8\pi(P_e - P_i)}$.

The pressure stratification in the umbra is very close to that one determined in the model with volume currents (Jahn 1989). The same holds for the total gas pressure difference; including the local maximum at the Wilson depression (cf. $\Delta B_{tot}$), which must be balanced by the magnetic forces. That implies a particular variation of the surface currents with depth. A characteristic feature of all models with the total magnetic flux larger than about $5 \cdot 10^{21}$ Mx is that the intensity of currents on the peripatopause exhibits a pronounced minimum below the Wilson depression (see $\Delta B_t$ in Fig. 3). The width of this minimum increases with the depth of the penumbra, which becomes deeper for larger magnetic fluxes. However, for fluxes smaller than $5 \cdot 10^{21}$ Mx the vertical extent of the penumbra is so small that there is no room for such a minimum.

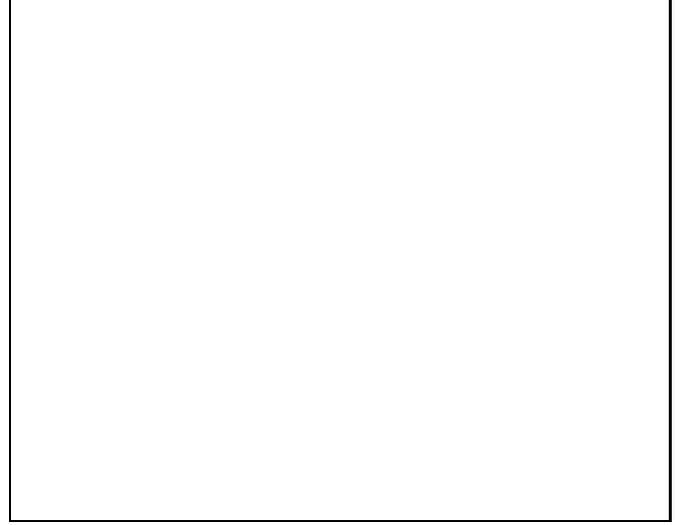

**Fig. 3.** Typical variation of the magnetic field jump with depth at the magnetopause ($B_t$), and at the peripatopause ($\Delta B_t$). The gas pressure difference between the solar convection zone and the umbra is plotted by means of $\Delta B_{tot} = \sqrt{8\pi(P_e - P_i)}$. Vertical dotted lines mark: the solar photosphere, the Wilson depression of the umbra, and the bottom of the penumbra

Thus, in models of larger spots the peripatopause contains two 'rings' of intense electric currents: one just above the bottom of the penumbra, and the second near the Wilson depression of the umbra. The lower 'ring' contributes only a little to the force balance, since its currents have an intensity almost an order of magnitude smaller than the intensity of currents at the magnetopause. The upper 'ring' plays a more important role. Here, the intensity of currents on both current sheets is comparable so that the gas pressure deficit in the umbra is balanced in part by the peripatopause and in part by the magnetopause. As a result, the configuration of the magnetic field is free of strong constrictions which might appear due to a local maximum of the total gas pressure difference (see Sect. 3.3 in Jahn 1992). The detailed distribution of currents and thereby the overall geometry of the spot model depends on all its free parameters, which are discussed below.

### 4.1. Parameter study

Besides the obvious dependence of $R_p$ on $\Phi$, the surface signature of the models is not altered very much by changes of the free parameters, because it has been fitted to the features of a 'typical spot' (cf. Sect. 3.5). Only the surface profile of the magnetic field may vary slightly from model to model, since it reflects the global geometry of the flux tube. The main differences among the models presented below are those related to the structure of the flux tube deep beneath the photosphere. This is, for instance, the depth of the penumbra $z_{bp}$, which changes mainly with the amount of the heat flux transmitted through the

### 4.1.1. Position of the bottom of the model

The base of the spot has been defined as the depth at which the magnetic field becomes purely vertical. The compression of the heat flux should then depend on the level $z_b$ chosen. However, such a dependence does not appear monotonous. That allows us to define a range of depths $z_b$ for which the models are physically plausible.

First, let us discuss the effect of the position of the bottom on the overall structure of the spot model. A comparison for two exemplary solutions for different values $z_b$ is given in Fig. 4. Both models have the same total magnetic flux, $\Phi = 10^{22}$ Mx, and the same transmissivity coefficient, $\epsilon = 0.7$. The position of the spot's bottom in the model **a** has been chosen at the depth $z_b = -15$ Mm, and in model **b** at $z_b = -20$ Mm. The shading represents the variation with depth of the heat flux ($\mathcal{F}$) and of the field strength ($\mathcal{B}$) inside the flux tube. The horizontal distance between lines is proportional to the value of the heat flux (darker regions have lower $F(z)$), and inversely proportional to the field strength. Beside the basic free parameters $\Phi$ and $\epsilon$, the fractional magnetic flux of the penumbra $\phi_{\mathrm{pen}}$ is printed in the figure, and the position of the penumbral bottom $z_{\mathrm{bp}}$ is indicated by the arrow. The surface profile of the magnetic field is drawn for each model in the lower right corner of the plots, and compared with the profile of Beckers & Schröter (1969) represented by the dotted line. The profiles are normalized to the central field strength and extend to the radius of the spot, which is calculated either at the Wilson depression of the penumbra (dashed line) or at the photosphere of the quiet Sun (continuous line).

Both models are very similar in the upper portion of the flux tube: they have the same surface radius $R_p \approx -14$ Mm, and the depths $z_{\mathrm{bp}}$ of both penumbrae differ only by a few hundred kilometers. However, the deeper solution ($z_b = -20$ Mm) is slightly narrower than the shallow model ($z_b = -15$ Mm) already at depths of about $-4$ to $-4.5$ Mm. The difference between the lateral extensions of the flux tubes increases with depth, so that the deeper model is more compressed at large depths, and has there values of the heat flux and of the field strength systematically larger. Moreover, the fraction of the magnetic flux contained in the penumbra $\phi_{\mathrm{pen}}$ is slightly smaller in the deep model, so that the field strength in the penumbra is systematically smaller than in the model with $z_b = -15$ Mm, see the magnetic profiles in Fig. 4.

Thus, the main differences among the models with various $z_b$ are those related to slightly different overall geometry, which affects the gross channelling effect of the heat flux. A simple and convenient measure of the degree of channelling in the model is the value of the compressed heat flux at the bottom $F_{\mathrm{ub}}$. The dependence of $F_{\mathrm{ub}}/F_\odot$ on the position of the bottom $z_b$ is plotted in Fig. 5. Each continuous line represents a series

**Fig. 4.** An analogue representation of two spot models differing only by the position of the bottom $z_b$: **a** $-15$ Mm, **b** $-20$ Mm. Values of the total magnetic flux $\Phi$, the fractional penumbral flux $\phi_{\mathrm{pen}}$, and the transmissivity of the magnetopause $\epsilon$ are printed for each model. The bottom of the penumbra is indicated by an arrow, and the surface profile of the magnetic field is shown in the right corner. Dashed and continuous lines correspond to the profiles normalized to the central field strength and extended to the radius of a spot as measured at the Wilson depression of the penumbra and at the solar photosphere, respectively. The dotted line represents the profile of Beckers & Schröter (1969). The horizontal distance between the shading lines on the left (right) side is inversely proportional (proportional) to the field strength (heat flux). In part **a** the region of the flux tube plotted below $z_b = -15$ Mm is extrapolated by the plotting routine

of spot models with a fixed value of the total magnetic flux $\Phi$. The transmissivity parameter $\epsilon = 0.7$ has been assumed in this series. For smaller values of $\epsilon$ the curves would be shifted slightly upwards (cf. Sect. 4.1.2.).

An interesting feature of our model is the minimum of the compression factor at $z_b = -15$ Mm. On both sides of this critical level there exist formally solutions that require a compression of the heat flux to values larger than the value of the solar heat flux ($F_{\mathrm{ub}} > F_\odot$). Such unrealistic models must be rejected. Plausible models (with $F_{\mathrm{ub}} \leq F_\odot$) can be constructed only for a limited range of $z_b$, which depends on the total magnetic flux of a spot: it is smaller for smaller $\Phi$'s. The model

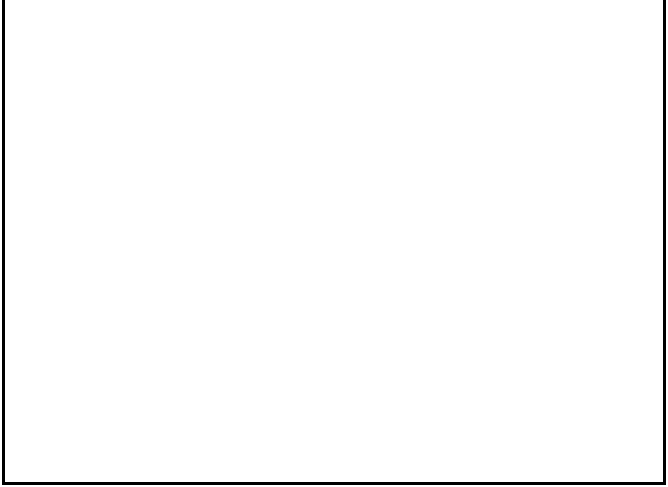

**Fig. 5.** Compressed heat flux $F_{ub}$ at the bottom $z_b$ in four series of models parametrized by different total magnetic fluxes (see labeling of continuous curves). The transmissivity coefficient $\epsilon = 0.7$ in all models. The minimum of $F_{ub}$ in each series coincides with the maximum ability of the matter to transport the heat by convection, as represented by the ratio of the convective to the radiative diffusivity $\kappa_{con}/\kappa_{rad}$ (dashed line). The scale of $\kappa_{con}/\kappa_{rad}$ (right abscissa) is normalized by the maximum value of this ratio. See also Appendix B and Fig. B2

with $\Phi = 10^{22}$ Mx, for example, may have the bottom placed at a depth in the range from $-20$ to about $-12$ Mm, whereas the solutions with $\Phi = 10^{21}$ Mx do not yield any reasonable models (see Fig. 5).

The characteristic shape of the dependence $F_{ub}(z_b)$, with the minimum at $z_b \approx -15$ Mm, reflects the partial ionization effects on the ability of the plasma to transport the energy by convection. This ability is described by the ratio of the convective to the radiative conductivity $\kappa_{con}/\kappa_{rad}$. Note, however, that although this ratio is related to the convective efficiency, it is not its direct measure unless the actual temperature gradient or the heat flux is specified (see Appendix B for details). The dashed curve in Fig. 5 shows the variations with depth of the ratio $\kappa_{con}/\kappa_{rad}$ (normalized to the maximum value) in the solar convection zone, with $\kappa_{con}$ calculated within the framework of the mixing length theory. The local maximum of $\kappa_{con}/\kappa_{rad}$ at the depth of about $-14.5$ Mm is related to the second helium ionization zone. The heat capacity of convective elements is enhanced in the region of partial ionization which increases the ability of matter to convect energy. Therefore the ratio $\kappa_{con}/\kappa_{rad}$ reaches a local maximum in layers where about half the chemical element (here He II) is ionized (see, e.g. Cox & Giuli 1968, Chap. 14). In our solar convection zone model that occurs at $z \approx -14.5$ Mm. Partial ionization of hydrogen and of He I has a similar, though somewhat smaller effect on $\kappa_{con}/\kappa_{rad}$ in shallower layers (see Fig. 5). However, our present models do not have bottoms placed higher than $z_b \approx -12$ Mm because we encounter convergence problems in the 'external' iteration (cf. Sect. 3.5.) which adjusts the thermodynamic stratifications to the magnetostatic equilibrium. This may be related to the lower boundary condition which would require then a purely vertical magnetic field at depths most probably too small.

The conductivities ratio inside the umbra near its base exhibits similar variations with depth as in the quiet Sun, since the stratification of the interior is very similar to that in the exterior at large depths. Gas densities are the same there and the internal gas pressure and the temperature gradient differ from the external ones by a factor $(1 - \beta^{-1})$. For the typical values of $B_b$, of the order of $10^4$ gauss, the plasma beta is equal to about $10^3$ or $10^4$ depending on the depth. Then, the ratio of $\kappa_{con}/\kappa_{rad}$ calculated near the bottom of the spot model has basically the same shape as the one in Fig. 5, but shifted slightly downwards by the depth of the Wilson depression.

The umbral luminosity ($\pi R_u^2 F_u$) is constant with depth. Larger convective ability at the bottom of the spot, allows for smaller divergence of the heat flux, which must fit the value $F_u$ at the surface. If we choose the maximum $\kappa_{con}/\kappa_{rad}$, i.e. $z_b = -15$ Mm, then in the spot models with $\Phi = 10^{22}$ Mx, for instance, only a threefold compression of the heat flux is required in order to fit the surface umbral flux. If we move the base into the deeper or shallower layers, then the heat transported by convection will be reduced, because of the lower convective ability. This reduction can be compensated only by an increased compression of the heat flux at this level.

The above reasoning can be applied if the convective efficiency (as measured by the mixing length parameter, for instance) is the same in all models. The same efficiency of convection at various positions of the spot's bottom can be expected, because the thermodynamic state of the base is, roughly speaking, determined by the scaling of the external stratification. The value of $\alpha$ in the solar convection zone is assumed to be independent of depth. The same feature should be exhibited by the bottom layer of the spot. Indeed, values of $\alpha_{bu}$ obtained in all models lie in the range from 0.68 to 0.79. The differences among them are due to the limited accuracy of the iteration procedure, and one can argue that $\alpha_{bu}$ is independent of $z_b$, $\Phi$, or $\epsilon$. Thus, the dependences $F_{ub}(z_b)$ shown in Fig. 5 follow in fact the curves of constant parameters $\alpha$ (cf. Fig. B3).

Summing up, the model discussed allows us to indicate a range of depths at which the flux tube constituting a spot might become a purely vertical cylinder. For the largest spots (of $\Phi = 10^{22}$ Mx) this can occur in the range from about $-20$ to $-12$ Mm, and for small ones ($\Phi \approx 5 \times 10^{21}$ Mx) only from $-17$ to $-13$ Mm. For the further discussion we make a conservative choice and fix the position of the bottom at the value $z_b = -15$ Mm, so that we choose the smallest possible divergence of the heat flux.

4.1.2. Transmissivity of the magnetopause

The above discussion of the convective ability and its effect on the channelling ignores the fact that the luminosity of the whole spot model may vary proportionally to the amount of energy entering the penumbra through the magnetopause (cf. Sect. 3.3.). This additional flux, parametrized by the transmissivity $\epsilon$, has an effect on the stratification of the gas in the penumbra model, and thereby on its depth $z_{bp}$ and also on the shape of the peripatopause. Therefore, the compression of the heat flux $F_{ub}$ at the base may depend also on the transmissivity. This dependence, for a series of models with three different values of the total magnetic flux, is shown in Fig. 6. Two models for transmission coefficients $\epsilon = 1$ and $\epsilon = 0.65$ and other parameters being equal are shown in Fig. 7.

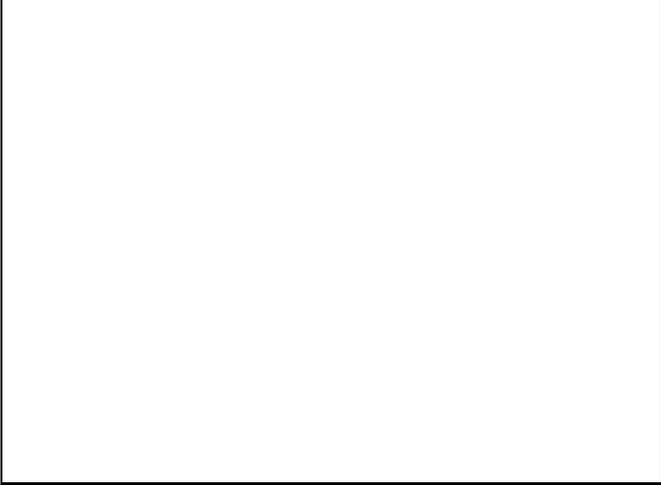

**Fig. 6.** Compressed heat flux $F_{ub}$ at the bottom ($z_b = -15$ Mm) as a function of the transmissivity coefficient $\epsilon$ in three series of models parametrized by different values of the magnetic flux (see labeling of curves)

When the transmissivity of the magnetopause becomes very large, i.e. larger than about 70%, then the compressed flux $F_{ub}$ remains practically independent of $\epsilon$. The main effect of the increase of $\epsilon$ is the reduction of the area of the permeable part of the magnetopause above $z_{bp}$, so that the total energy input does not exceed the total irradiance of the penumbral photosphere. The necessary reduction can be achieved by a relatively small change of $z_{bp}$, because of the large inclination of the magnetopause at those depths. For this reason, the gross channelling effect on $F_{ub}$ together with the overall mechanical equilibrium, remains unaffected for large $\epsilon$. This is a situation, in which the penumbra above $z_{bp}$ forms a relatively shallow layer (cf. Fig. 7a). Then, the stratification in the penumbra is closer to that in the quiet Sun than to the umbral one. Consequently the electric currents are more intense on the peripatopause than on the magnetopause and the inner current sheet plays a dominant role in the lateral force balance between the umbra and the unperturbed convection zone at the relevant depths. As a result there would evolve a growing jump in the photospheric field strength between penumbra and umbra (cf. Fig. 7a), which is not observed. Therefore we must assume $\epsilon \lesssim 0.7$.

For lower transmissivity coefficients less energy enters the penumbra, so that its pressure stratification becomes more different from the external one and closer to the umbral pressure distribution. Consequently, the contribution of the peripatopause to the lateral force balance decreases, since the related electric currents become less intense. The overall magnetostatic equilibrium is dominated by the currents confined to the mag-netopause. The umbra must then have a more tapering shape, roughly such as implied by the shape of the magnetopause. Eventually, the divergence of the heat flux becomes more pronounced, so that the compressed heat flux $F_{ub}$ increases (see Fig. 6).

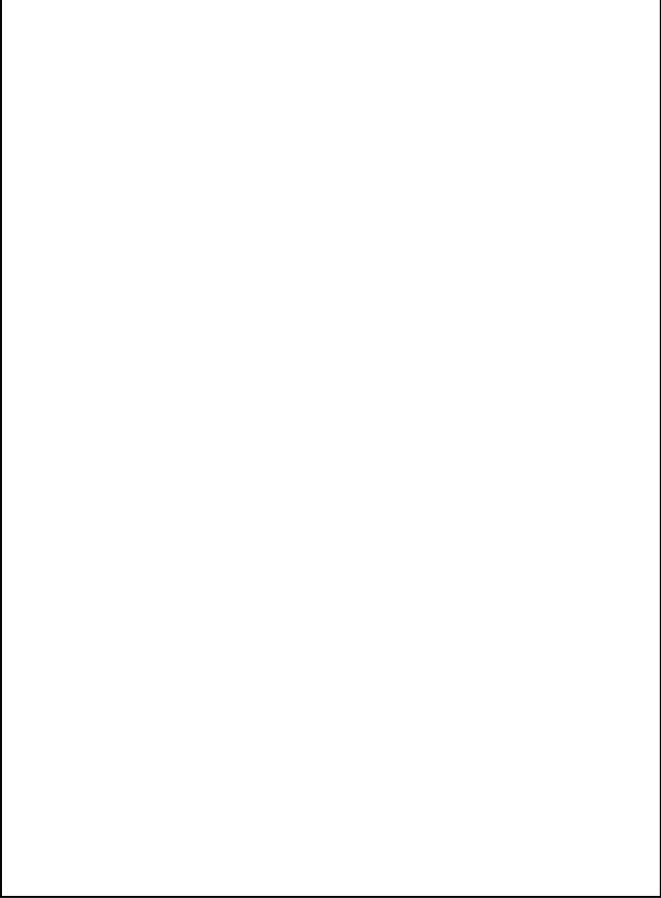

**Fig. 7.** The effect of the transmissivity coefficient $\epsilon$ on the structure of the penumbra: a) $\epsilon = 1$, b) $\epsilon = 0.65$ (see also Fig. 8c for the intermediate value $\epsilon = 0.7$). The models are represented in the same form as in the Fig. 4. The bottom of both models is placed at $z_b = -15$ Mm. Note the large inclination of the magnetopause far below the bottom of the penumbra in the model with a fully free inflow of energy ($\epsilon = 1$)

The transmissivity of the magnetopause must not be too small for two reasons. First, the heat compression increases with the decreasing $\epsilon$ and the models enter the region of unrealistic solutions (with $F_{ub} > F_\odot$) when the transmissivity is too low. For the magnetic magnetic flux $\Phi = 10^{22}$ Mx, for example, such fictitious solutions are obtained if the assumed transmissivity is smaller than about 55%. Slightly smaller values of $\epsilon$ can be reached in models of larger spots, but then another limitation appears. The tripartite model requires some minimum heat supply from outside. If the magnetopause transmits less than about 50% of the solar heat flux, then the energy supply is too small to construct any model that has a properly extended penumbra with the prescribed heat flux at the surface of $0.75\ F_\odot$.

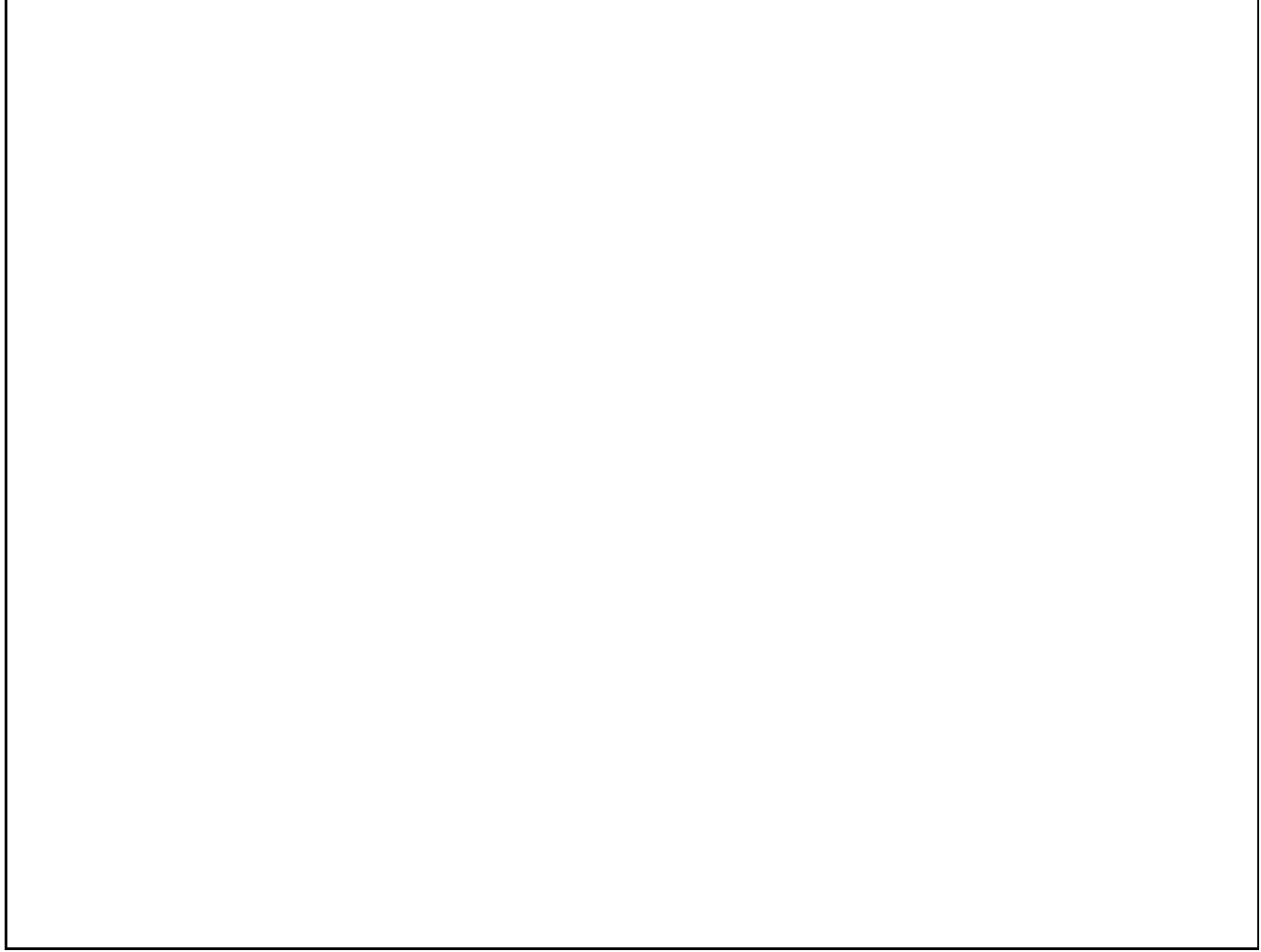

**Fig. 8.** Four models with different values of the total magnetic flux in the same representation as in the Fig. 4. Model a is classified as an unrealistic solution, for the compressed heat flux $F_{ub}$ is larger than $F_\odot$. Values of $F_{ub}/F_\odot$ are the following: **a** - 1.23, **b** - 0.93, **c** - 0.79, **d** - 0.62. The frames are foreshortened horizontally by a constant. **d** is demagnified by 20 percent.

Although there are no strict criteria according to which a particular value of the transmissivity coefficient can be chosen, we conclude from the preceeding that the range of feasible values is such that $\epsilon \approx 0.6 - 0.7$. This argument is corroborated by the conjecture that the interchange convection operates effectively only at depths at which the inclination of the magnetopause to the vertical exceeds some minimum value between 20 and 30 degrees. Inclination angles in this range are found indeed in models with the transmissivity coefficients $\epsilon$ in the range from 0.6 to 0.7. In such models the penumbra merges with the umbra 5 to 6 Mm beneath the surface, where the magnetic field becomes more vertical.

4.1.3. The total magnetic flux of a spot

Compression of the heat flux $F_{ub}$ is larger in models with smaller magnetic flux $\Phi$ (cf. Fig. 5). However, the structure of upper parts of the spot, including the whole penumbra, does not depend on the horizontal size of the flux tube. The thermal and magnetic state of the penumbra and of the adjacent part of the umbra is similar in all models with a magnetic flux larger than a certain critical value.

Four models presented in the Fig. 8 are computed with the same transmissivity coefficient $\epsilon = 0.7$, the same position of the bottom $z_b = -15$ Mm, and for four different values of the total magnetic flux: $10^{21}$, $5 \cdot 10^{21}$, $10^{22}$, and $2 \cdot 10^{22}$ Mx. The smallest flux tube (model **a**) has the surface radius $R_p$ of about 5 Mm, while the largest one (model **d**) has $R_p \approx 20$ Mm. Model **a** has the largest compression of the heat flux ($F_{ub} = 1.23 F_\odot$) and is classified as an unrealistic solution. In the remaining models values of $F_{ub}$ are equal to 0.93, 0.79, and 0.62 $F_\odot$, in the order of the increasing total magnetic flux $\Phi$.

It is remarkable that all models have a similar geometry of the upper region extending from the surface down to the merging level $z_{bp}$. This similarity relates to the fact that the energy irradiated by the penumbra at the surface consists mostly of the heat that enters through the magnetopause. If we ne-glect the small contribution of the heat flux coming from below the penumbra's bottom, then its surface luminosity should be equal to the total energy crossing the magnetopause per unit time. For the transmissivity coefficient $\epsilon$ fixed, this energy is proportional to the projected area of the permeable part of the magnetopause, or the 'transmission surface'. Also the surface luminosity of the penumbra depends only on its area, because the heat flux is fixed. Since the total heat acquired through the magnetopause is equal to the total heat lost at the surface, the ratio of the penumbral area to the projected area of the 'transmission surface' is also fixed and should be independent of $\Phi$. When the magnetic flux is increased, then both areas must increase proportionally to the radius squared. As a result the 'transmission surface' must increase also its vertical extent: $z_{bp}$ is shifted downwards. Since $z_{bp}$ grows proportional to the radius our models are geometrically similar and the heat flux in the umbra at the merging level $z_{bp}$ is compressed to almost the same value in all models: the value $F_u(z_{bp}) = F_p(z_{bp})$ is equal to about $0.4 F_\odot$ with an accuracy of about 5%.

The further compression of the heat flux, below $z_{bp}$, does depend on the available range $z_{bp} - z_b$, i.e. on $z_{bp}$ and thereby on the total magnetic flux. The latter dependence is presented in the Fig. 9 for the transmissivity parameter $\epsilon$ in the range between 0.6 and 0.7, and can be used to determine the range of magnetic fluxes for which plausible models of spots can be constructed. Note, that for the conservative choice of the position of the bottom $z_b = -15$ Mm all such models represent the minimum possible compression of the heat flux. For any other choice of $z_b$ the curves $F_{ub}(\Phi)$ in the Fig. 9 would be shifted upwards.

As a result of the larger available range $z_{bp} - z_b$ the compression of the heat flux at the bottom is larger in smaller spots and the value of $F_{ub}$ may exceed $F_\odot$ if the magnetic flux is too small. The smallest plausible spot model, with the marginal value of $F_{ub} = F_\odot$, has a total magnetic flux of about $3 \cdot 10^{21}$ Mx (cf. Fig. 9). For still smaller values of the magnetic fluxes the solutions with a properly extended penumbra require more energy at the base than actually is available (e.g. model **a** in the Fig. 8). Since we do not study spot models of different field strengths or effective temperatures, the minimum value of $\Phi$ derived here is not an absolute limit for all spots. If we would fit the model to a smaller field strength, for instance, then its radial size would be larger and the critical value of $\Phi$ should decrease according to the argumentation above. Precise determination of such a lower limit requires, however, much more extensive study then the present one. Our result indicates only that there exists a critical value of the total flux $\Phi$, below which a magnetic field concentration modelled cannot form a regular stable sunspot with a penumbra.

There are some indications, however, that models of pores with the compressed flux $F_{ub} \leq F_\odot$ might be constructed within the same general approach. The degree of compression decreases when the fraction $\phi_{pen}$ of the magnetic flux in the penumbra is decreased. Of course, such models do not fulfill the requirement that $R_u = 0.5 R_p$. The depth of such small 'penumbrae' also decreases: $z_{bp}$ is shifted upwards so that the magnetopause reaches the critical inclination at which the interchange convection becomes effective only in layers very close to the surface. This tendency indicates that in the limit $\phi_{pen} \rightarrow 0$ a pore model could be obtained with $F_{ub} \leq F_\odot$. It is tempting to relate this feature to the transition from a sunspot to a pore. The inverse transition, from a pore to a spot, was considered by Simon & Weiss (1970) in discussing a model of a pore. Their model breaks down for the magnetic fluxes that are too large, because the boundary layer surrounding the flux tube was assumed to be infinitesimally thin. The lack of a finite thickness interface between the interior and the exterior causes that very large pore models must have a central surface field strength of more than 3500 gauss in order to balance the external gas pressure as was confirmed by Schmidt and Wegmann (1983). Our present models always include an extended boundary layer, which at the surface reaches the thickness of one half of the radius of the flux tube. These models break down for magnetic fluxes that are too small.

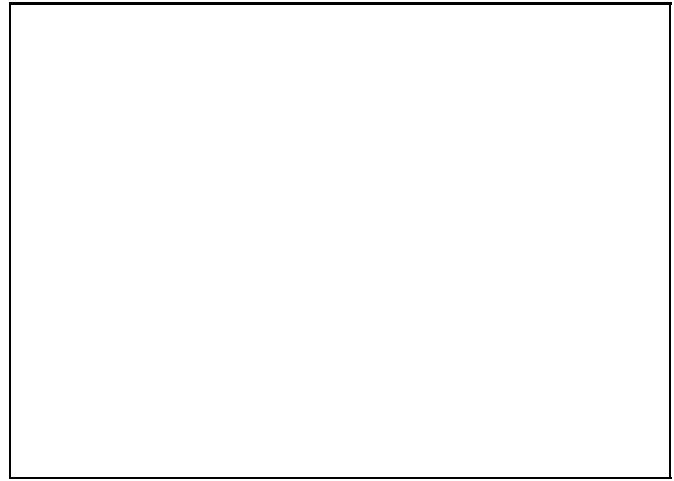

**Fig. 9.** Compressed heat flux at the bottom $F_{ub}$ as a function of the total magnetic flux $\Phi$. The two curves represent the range of the feasible values of the transmissivity coefficient $\epsilon$. Solutions with $\Phi < 3 \cdot 10^{21}$ Mx (for $\epsilon = 0.7$), and with $\Phi < 6 \cdot 10^{21}$ Mx ($\epsilon = 0.6$) are regarded as unrealistic, for the value of $F_{ub}$ exceeds $F_\odot$

### 4.2. Stability of models against fluting

The magnetopause of a sunspot is concave towards the field–free plasma, and might therefore be subject to the fluting instability (Parker, 1975), but the buoyant effects of the gravitational stratification of the magnetized plasma can counteract the fluting tendency. Meyer et al. (1977) investigated these effects using the energy principle formulated for a magnetic interface by Bernstein et al. (1958). The condition for stability they derived requires that the magnetic field fans out with height rapidly enough, so that the radial component of the magnetic field at the magnetopause decreases with height ($d|B_r|/dz < 0$). The condition can also be expressed as a criterion relating the cur-

vature forces and the buoyancy of the magnetized plasma (cf. Eq. (2.14) in Meyer et al. 1977):

$$\frac{B_t^2}{4\pi r_c} < g(\varrho_e - \varrho_i)\sin\theta, \tag{8}$$

where $r_c$ is the curvature radius, and $\theta$ is the inclination of the magnetopause. Other symbols have their usual meaning. According to this criterion the flux tube can be stable but only if the interior contains less dense matter. Then, the buoyancy wins over the curvature force which induces the fluting.

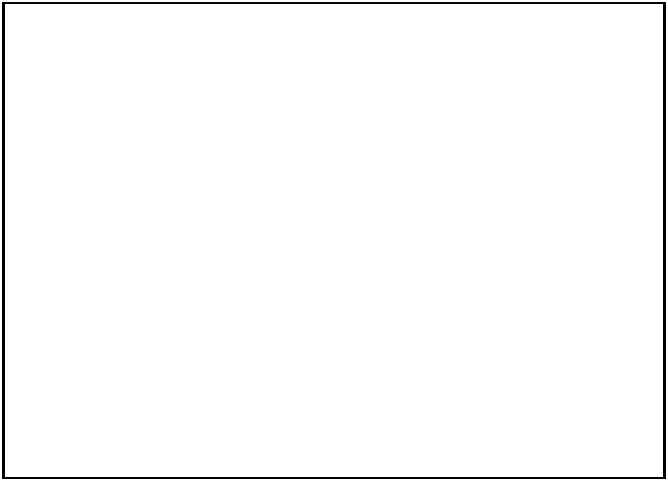

**Fig. 10.** Stability of the magnetopause in the exemplary model (cf. the model **b** in the Fig. 7), according to the criterion derived by Meyer et al. (1977). The narrow spike corresponds to the density inversion that appears near the surface in the penumbra model

Applying Meyer et al.'s condition to our model we find that the magnetopause is stable against fluting in the upper part of the flux tube. Figure 10 shows the stability criterion in the form of the vertical gradient of the radial field strength $B_r$ at the magnetopause in the model presented in the Fig. 7b. The local variance and instability in a shallow layer near the surface of the penumbra is induced by the density inversion that appears in convective envelope models of late type stars, i.e. in both stratifications, the quiet sun and at a slightly lower level the penumbra. We consider this feature as inessential because except for a very small region the curvature forces are sufficiently counteracted by the buoyancy of the penumbra where the density is systematically smaller than in the quiet sun.

Stability near the base of the spot model is marginal. Our lower boundary conditions require that both the density difference between the exterior and the interior, and the curvature force tend to zero in a similar manner. Therefore, the differences between the two forces become very small. Because of the limited accuracy of numerical procedures, we cannot decide with confidence whether the lowest part of the flux tube (see Fig. 10) is marginally stable or unstable.

At intermediate depths between 5–11 Mm, the density inside the tube model becomes comparable to that outside, and buoyancy becomes indeed slightly too weak to compensate for the remaining curvature forces. Therefore the magnetopause becomes unstable in this range. It is not clear, however, how far and how fast this instability can proceed, and whether the nonlinear effects may not lead to a saturation of the process, so that a kind of diffuse magnetopause may be formed. The layer at which the instability appears coincides with the depth at which the penumbra merges with the umbra in our model. It is conceivable that the parametrized transmissivity of the magnetopause is related to the fluting instability. The fastest growing modes are those with the shortest wavelengths, according to a rough estimate by Bernstein et al. (1958). Perhaps those modes could make the magnetopause diffuse and thereby resistant to more fluting. They may also increase its transmissivity to heat flux up to the estimated level of 60–70%, which might not be achieved by a pure radiative exchange.

## 5. Conclusions

The internal constitution of sunspots can be investigated by means of a global approach, in which some observed signatures of spots are regarded as a surface manifestation of the overall configuration of the magnetic flux concentrations in the solar convection zone. Such an approach is corroborated by the existence of some well defined typical values of spot observables like the heat flux in umbra and penumbra, the radial magnetic profile, and the ratio of the umbral and penumbral area. We adapted these features as constraints imposed on the photospheric layers of a magnetostatic flux tube model, which simulates the global structure of an isolated sunspot. The model represents the upper part of the magnetic flux concentration in the solar convection zone, where the assumption of a global magnetostatic equilibrium is justified. As in previous investigations of that type (Schmidt & Wegmann 1983, Jahn 1989) we use the free boundary value formalism to relate the surface constraints to the magnetic field geometry deep below the photosphere. The assumption of a perfect thermal isolation of the umbra and a partial one of the penumbra, and the approximation of both regions by 1-D stratifications enables us to obtain a fully self-consistent coupling of the magnetohydrostatic equilibrium with the thermodynamics of the spot's interior including the heat flow.

The assumptions we made lead to the following picture of a sunspot. Deep below the surface, where the magnetic field is vertical, the entire flux tube is thermally isolated from the surrounding convection zone. Here an almost vertical magnetopause implies purely radiative heat exchange between the exterior and the interior, which is neglected for the time scales considered. The energy inside the flux tube is transported by some magnetoconvective process of smaller efficiency as compared to the quiet Sun, and the heat flux is diluted as the tube widens with height. Above a certain depth the inclination of the magnetic field in the outer part of the tube, the penumbra, is large enough for interchange convection to set in. Due to some local instabilities the heat transport across the inclined magnetopause becomes so efficient, that a significant part of the impinging solar heat flux can be transmitted and carried upwards and inwards by the interchange convection. The range of inward penetration of alternating magnetic flux elements is limited by the inclination of the magnetic field. The motion ceases at the peripatopause, where this inclination is too small. There is no further efficient heat transport inwards, so that the luminosity of the umbra is still constant with depth. The penumbra forms a horizontally stratified layer due to an efficient lateral mixing and increases its luminosity with height as it is fed with energy from outside. We impose the features of the 'typical sunspot' at the surface of the flux tube, i.e. the appropriate values of the heat fluxes, of the field strength, and of the relative extension of the penumbra. Thus, the possible overall structure of a spot discussed below is implied by these constraints as well as by the two main assumptions of the model: the thermal isolation of vertical magnetic flux and the interchange convection in the inclined penumbra.

Due to the divergence of the heat flux with height this tripartite model has the convection at large depths at least three times more efficient than the previous model without channelling (Jahn 1989). It is even possible to construct models exclusively within the scope of Hoyle's (1949) concept, i.e. with a heat flux at the base equal to the normal solar flux, so that the efficiency of convection is the same as in the unperturbed surroundings. We do not find, however, any indication that some particular efficiency near the base is preferred: models without any suppression of convection, and models with an efficiency reduced to about 70% of normal solar convection are indistinguishable at the surface.

Thus, the present model does not encounter an unavoidable problem of heat pile-up which might quickly destruct the spot (e.g. Parker 1974, 1976). That refers not only to the base of the spot, but also to relatively shallow layers, of about 2000 km beneath the surface, where the convective efficiency in the magnetic field decreases with height (Meyer et al. 1974). If channelling were excluded, then one would have to assume an efficient heat exchange with some other form of energy transport in order to remove the heat excess that cannot be conveyed vertically by an inefficient oscillatory convection. Such an implicit assumption was made in the previous model (Jahn 1989) for the layers where the convective efficiency measured by the mixing length parameter decreased abruptly with height (cf. also Eq. (3)). A similar decrease of the efficiency is compensated in the present model by the decrease of the vertical heat flux due to the channelling, so that no heat removal must be assumed.

The stratification of the umbra in the tripartite model exhibits a similar local deficit of the gas pressure at the level of the Wilson depression as in the previous model (Jahn, 1989). It is difficult to ascribe this feature to inadequate assumptions made in magnetostatic models, since it appears also in numerical simulations of umbral structure by Nordlund & Stein (1990). It leads to an excess of the horizontal gas pressure difference $\Delta P$, which should be balanced by magnetic forces in such a way that the magnetostatic configuration of field lines is free of strong unstable constrictions. In this fashion a penumbra with interchange convection seems to accommodate the excess. One can speculate that this excess might even initiate the formation of the penumbra in magnetic flux tubes originally isolated thermally from the surroundings. Then, the only current sheet yet existing, the magnetopause, would locally bend inwards at the level of the Wilson depression. This would make the magnetic configuration at this level prone to interchange instability which might then initiate the formation of a penumbra. Such a scenario concurs with observations of Bumba and Suda (1984), who observed the formation of the penumbra. The process started at that site of the umbra at which the radial gradient of the magnetic field was the steepest. It would correspond to the region of the constricted magnetic field due to the inward bend of the magnetopause. The umbral field intruded the photosphere forming locally a few filaments. This process continued successively around the umbra and was completed after a few hours.

Our model describes the situation in which the magnetic field in the outermost part of the tube found an equilibrium state without strong constrictions by forming a thick penumbra with interchange convection. This region separating the umbra from the quiet Sun contains 50-60% of the total magnetic flux of a spot. The inclination of the magnetic field in this region, of more than about 30°, is apparently sufficient to maintain the interchange convection. Such a penumbra allows the inflow of heat through the magnetopause and redistributes the acquired energy over large horizontal distances. The brightness of the penumbral photosphere and its radial extent require that the permeable part of the magnetopause transmits about 60% of the impinging solar heat flux and extends down to a depth comparable with half of the surface radius of the spot. This self–similar geometrical structure keeps the total flux divergence in the adjacent umbra almost constant. It appears that a penumbra which fulfills these conditions can reduce the remaining flux divergence or channelling in the umbra, down to the bottom level of the model, sufficiently only in spot models with the magnetic flux larger than about $3 \cdot 10^{21}$ Mx. Otherwise the compression of the heat flux at the base is too large so that such models break down.

Our magnetostatic model for a sunspot includes the approximation for the thick penumbra based on the concept of the interchange convection. It accounts for the observed characteristic global features of spots and gives some interesting conjectures concerning their structure at large depths. However, farther theoretical study of the dynamics of the penumbra is required in order to discuss quantitatively the physical mechanisms related to the efficiency of the interchange convection, the possible range of its inward penetration, the size and shape of individual convective flux sheets, or the heat transport through the magnetopause.

*Acknowledgements.* We are indebted to Rudolf Wegmann for the enlightening discussions on the free surface problem in the potential theory. KJ acknowledges the hospitality and the support from the Max Planck Institut für Astrophysik in Garching.

This work was also supported in part by the KBN grant 2-1213-91-01.

## Appendix A

This section is intended to present an extension of Wegmann's study of the free boundary problem of potential theory (Wegmann 1981, 1987) for our particular problem. Therefore, the formulae given below do not have a general shape but their notation is as close as possible to the one we solve numerically. The method we use basically follows the study of Schmidt & Wegmann (1983).

The approximation for the magnetic field of a sunspot is presented in Fig. A1. The magnetic domain considered is enclosed by the surface $\Gamma_1$ which is defined by the magnetopause and by the lower and the upper limiting surfaces. The second current sheet $\Gamma_2$, the peripatopause, is a magnetic surface placed inside the flux tube. The continuous line in the figure represents the part of $\Gamma_2$ which contains the surface currents. The magnetic field is normal on the upper and the lower boundaries and tangential on the magnetopause and on the peripatopause. Currents are distributed on $\Gamma_1$ and $\Gamma_2$, and their density is related to the jumps of the magnetic field at the respective surfaces. The potential $u$, which generates the magnetic field ($\boldsymbol{B} = -\nabla u$), is the sum of two contributions: $u_1$ – the potential from $\Gamma_1$, and $u_2$ – the one from $\Gamma_2$. These potentials can be calculated by means of Green's formula at each point excluding the surface containing the currents. There, the potentials $u_1$ or $u_2$ and their derivatives (that is the respective components of the magnetic field vector) must obey jump relations which yield equations for the determination of the free boundaries, i.e. the shape of the magnetopause and of the peripatopause.

As the first condition we use that the magnetic field outside the considered domain must be equal to zero. This requirement can be reduced to the problem with a single current sheet considered by Schmidt and Wegmann (1983), if we subtract the potential $u_2$ generated by $\Gamma_2$ from the total potential $u$. Then the flux tube contains no currents (sources) inside and the problem is basically the same, with the exception that the integral equation for the free surface $\Gamma_1$ reads now:

$$u_1^- = -u_2 \qquad (A1)$$

where the superscript minus denotes the boundary values of the potential $u_1$ outside of $\Gamma_1$ (cf. Eq. (2.3), Schmidt & Wegmann 1983). The potential $u_2$ and its derivatives are continuous across $\Gamma_1$ so that it can be determined by means of Green's formula, whereas in the evaluation of $u_1^-$ one has to take Green's formula for the boundary values.

Applying the general considerations of Wegmann (1981, 1987) for our particular problem we derive the following relations. The potential $u_1^-$ at the point $p$ of $\Gamma_1$ is given by the integral

$$u_1^-(p) = \frac{1}{2\pi} \int_{\Gamma_1} \left[ \frac{B_n(s)}{\rho} + \left(u_1^+(p) - u_1^+(s)\right) \frac{\partial}{\partial n} \frac{1}{\rho} \right] d\mathcal{O}(s) \quad (A2)$$

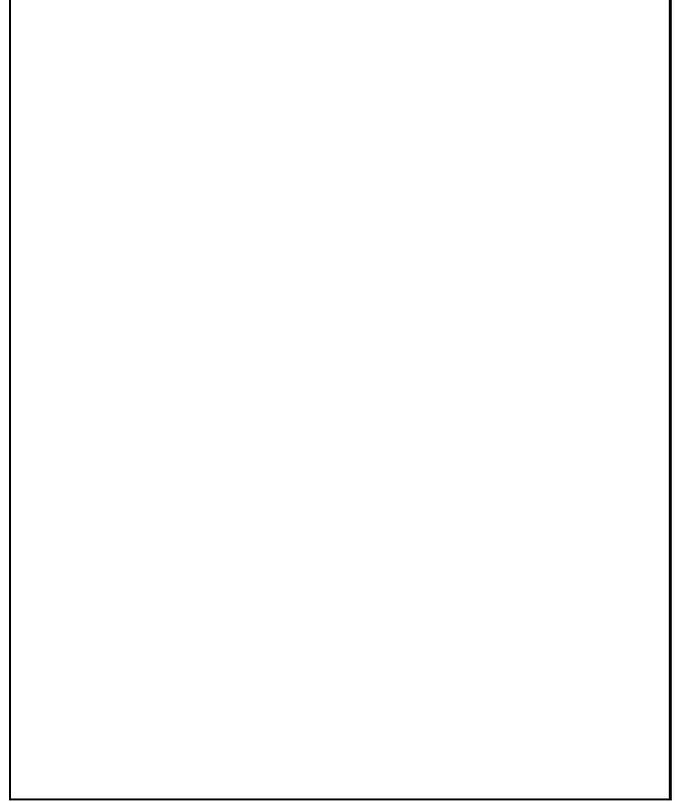

**Fig. A1.** Schematic illustration of the free surface problem for two current sheets. The surface $\Gamma_1$ bounds the domain considered and consists of the lower plane, the magnetopause, and the upper spherical cup. $\Gamma_2$ represents the peripatopause. The continuous line, extending between $z_{\mathrm{ib}}$ and $z_{\mathrm{it}}$, indicates the part of the peripatopause that carries the surface currents, while the dotted one shows its continuation along the magnetic field. $B_{1\mathrm{n}}$ and $B_{2\mathrm{n}}$ are the normal components of the magnetic field, while $u_1$ and $u_2$ are the scalar potentials, generated by $\Gamma_1$ and $\Gamma_2$, respectively

where $d\mathcal{O}(s)$ is the surface element, $\rho$ denotes the distance between the actual point $p$ and the source point $s$, and $-B_\mathrm{n}$ is the normal component of the field strength equal to the jump of the normal derivative of $u_1$. The boundary value of the potential from inside, $u_1^+$, is constant on the lower and on the upper surface. If $u_1^+ = 0$ on the lower surface is assumed then the potential from inside at the magnetopause can be determined by simple integration (cf. Eq. (2.1), Schmidt & Wegmann 1983):

$$u_1^+(s) = \int_0^s B_\mathrm{t}\, dl, \qquad (A3)$$

where $B_\mathrm{t}$ is the tangential component of the field strength representing the density of the surface currents, and $l$ denotes the arc length. The integration starts at the lower end of the magnetopause due to the orientation of $\Gamma_1$. The potential $u_1^+$ on the upper cup has the value as given by Eq. (A3) at the upper end of the magnetopause.

In cylindrical coordinates the surface element is given by $d\mathcal{O} = R\sqrt{1 + (dR/dz)^2}\,d\varphi dz$ (or $d\mathcal{O} = r d\varphi dr$ for the lower limiting surface), where the curve parametrized by $R(z)$ describes a meridional cross-section of the magnetopause and the upper cup. The inclination of the curve with respect to the vertical is given by the angle $\theta$ ($dR/dz = \tan\theta$). The normal to $\Gamma_1$ has the components $\boldsymbol{n} = (\cos\theta\cos\varphi, \cos\theta\sin\varphi, -\sin\theta)$, and the distance $\rho$ can be expressed in the following form: $\rho^2 = (z - z')^2 + r^2 + r'^2 - 2rr'\cos(\varphi - \varphi')$, where prime is used to denote collocation points. Let $z_b$ ($z_t$) denote the lower (upper) end of the magnetopause. Now the potential in the point $(r', \varphi', z')$ given by the integral in Eq. (A2) can be written as a sum of three contributions: from the lower cup –

$$\frac{B_t(z_b)}{\pi} \int_0^{R(z_b)} r \frac{K(k)}{b} \, dr, \qquad (A4.1)$$

from the upper cup –

$$\frac{B_t(z_t)}{\pi} \int_{z_t}^{Z} R\sqrt{1 + \left(\frac{dR}{dz}\right)^2} \frac{K(k)}{b} \, dz, \qquad (A4.2)$$

and from the magnetopause –

$$\frac{1}{\pi} \int_{z_b}^{z_t} \left(u_1^+(z') - u_1^+(z)\right) R \times$$
$$\left\{ \frac{E(k)}{a^2 b}\left[(z - z')\frac{dR}{dz} - (R - r')\right] - \frac{2D(k)}{b^3} r' \right\} dz. \qquad (A4.3)$$

$K(k)$, $E(k)$, and $D(k)$ are complete elliptic integrals with $k^2 = 1 - (a/b)^2$, where $a^2 = (z - z')^2 + (r - r')^2$ and $b^2 = (z - z')^2 + (r + r')^2$. For reasons of continuity the normal component of the field on the lower boundary is equal to the value of the tangential field on the magnetopause at $z_b$ ($B_t(z_b)$) and on the upper cup equal to the respective value at $z_t$ ($B_t(z_t)$). $Z$ denotes the height of the upper cup for $r = 0$ (see Fig. A1).

Equations (A4.1)-(A4.3) constitute the left hand side of Eq. (A1). The right hand side, i.e. the potential from the peripatopause, is calculated in the point $(r', \varphi', z')$ on $\Gamma_1$ by means of Green's formula, which in our notation has the following form:

$$u_2 = \frac{1}{\pi} \int_{z_{ib}}^{z_{it}} \left[\int_0^{l(z)} \Delta B_t dl_i\right] R_i \times$$
$$\left\{ \frac{E(k)}{a^2 b}\left[(z - z')\frac{dR_i}{dz} - (R_i - r')\right] - \frac{2D(k)}{b^3} r' \right\} dz, \qquad (A5)$$

where the curve $R_i(z)$ describes the peripatopause. The first term in the integral describes the sources of the potential $u_2$ on $\Gamma_2$, which are determined as previously (see Eq. (A3)) by the integration of the jump of the magnetic field $\Delta B_t$ along the meridional cross-section of the peripatopause. Thus, Eqs. (A4)-(A5) determine the full form of the integral condition from which the shape of the magnetopause can be calculated.

On the peripatopause the condition must be fulfilled that the normal component of the magnetic field is equal to zero. That yields the second equation:

$$B_{1n} = -B_{2n}, \qquad (A6)$$

which should be solved simultaneously with the Eq. (A1) in our two free surfaces problem. Index 1 or 2 indicates the field generated by $\Gamma_1$ or by $\Gamma_2$. We use Eq. (7.10) of Wegmann's analysis (Wegmann 1987) to derive the respective components of the magnetic field on $\Gamma_2$. The normal component from $\Gamma_1$ has the following form:

$$B_{1n} = \frac{B_t(z_b)}{\pi}(z_b - z') \int_0^{R(z_b)} \frac{E(k)}{a^2 b} r \, dr \; +$$
$$\frac{1}{\pi} \int_{z_b}^{z_t} B_t R \left\{ \frac{E(k)}{a^2 b}\left[(r' - R)\frac{dR}{dz} + (z' - z)\right] - \right.$$
$$\left. \frac{2D(k)}{b^3}\left[r'\frac{dR}{dz} + (z' - z)\right] \right\} dz \; + \qquad (A7.1)$$
$$\frac{B_t(z_u)}{\pi} \int_{z_t}^{Z} R \left\{ \frac{E(k)}{a^2 b}\left[(r' - R) - (z' - z)\frac{dR}{dz}\right] + \right.$$
$$\left. \frac{2D(k)}{b^3} R \right\} dz.$$

The three terms in this equation correspond to the integration over the lower surface, the magnetopause, and the upper cup respectively. $R$ refers either to the magnetopause (the second term) or to the spherical cup (the third term). The remaining symbols have their usual meaning (see also Fig. A1). The second contribution of the normal field on the peripatopause, from the peripatopause itself, is determined by the following integral:

$$B_{2n} = \frac{1}{\pi} \int_{z_{ib}}^{z_{it}} \Delta B_t(z) R_i \times$$
$$\left\{ \frac{E(k)}{a^2 b}\left[(z' - z) + (r' - R_i)\frac{dR_i}{dz}\right] - \right. \qquad (A7.2)$$
$$\left. \frac{2D(k)}{b^3}\left[r'\frac{dR_i}{dz} + (z' - z)\right] \right\} dz.$$

We solve both the Eq. (A1) and the Eq. (A6) for $\Gamma_1$ and $\Gamma_2$ simultaneously by a Newton method. Corrections $\Delta r_j$ for the positions of the magnetopause and of the peripatopause are determined from the linearized equations calculated in each collocation point $x_i$. These points are distributed along the whole $\Gamma_1$ and $\Gamma_2$ including the parts that do not contain sources (dotted lines in Fig. A1). Curves describing the magnetopause and the peripatopause are represented by cubic splines defined by the position of points $r_j = R(z_j)$ for fixed depths $z_j$. The change of the potentials $u_1^-$ and $u_2$ and the change of $B_{1n}$ and $B_{2n}$ is computed in the collocation points for each perturbation of the position of the points $r_j$. Thus, the shape of $\Gamma_1$ and $\Gamma_2$ is corrected at each iteration step by the solution of the following system of equations for corrections $\Delta r_j$:

$$\begin{cases} \sum_{j=1}^{m} \delta u_{ij} \Delta r_j + u_i = 0, & i = 1, 2, \ldots n_1, \\ \sum_{j=1}^{m} \delta b_{ij} \Delta r_j + b_i = 0, & i = 1, 2, \ldots n_2. \end{cases} \quad (A8)$$

where $i$ and $j$ denote $n_1 + n_2$ collocation points and $m$ perturbation points, respectively, and

$$\begin{aligned} \delta u_{ij} &= \left.\frac{\partial u_1^-}{\partial r_j}\right|_{x_i} + \left.\frac{\partial u_2}{\partial r_j}\right|_{x_i}, & u_i &= u_1^-(x_i) + u_2(x_i); \\ \delta b_{ij} &= \left.\frac{\partial B_{1n}}{\partial r_j}\right|_{x_i} + \left.\frac{\partial B_{2n}}{\partial r_j}\right|_{x_i}, & b_i &= B_{1n}(x_i) + B_{2n}(x_i). \end{aligned} \quad (A9)$$

The parameters of the model are the following: the total magnetic flux of the tube ($\Phi$), which defines the radius of the magnetopause at $z_b$ if the value of the field strength at the bottom is known ($B_t(z_b) = B_b$ according to the thermodynamic computations), and the fraction $\phi_{pen}$ of the total magnetic flux constituting the penumbra, which defines the position of $\Gamma_2$ with respect to the magnetopause. The position of the lower and the upper end of the magnetopause, $z_b$ and $z_t$, are placed sufficiently deep (or high), as we believe, so that they do not influence the essential part of the magnetic configuration modelled. In all models $z_t = 800$ km is chosen, and the 'shallowest' spot models have a depth comparable with the surface radius, i.e. $z_b = -12$ Mm (see Table 1).

## Appendix B

Here we present some thermodynamic properties of the spot's stratification near the base of the present model, and their effect on the value of the compressed heat flux inside the flux tube.

The lower boundary conditions in the model seem to be quite rigid, when the channelling of the heat flux is considered. The extreme assumption of perfect thermal isolation of the flux tube, and the approximation of the magnetic field at the base as completely homogeneous leave only little freedom for the thermal structure of the tube at the depth $z_b$. It must be very close to that one in the exterior. There are no other forces but those related to the magnetic field jump on the magnetopause, that might alter the stratification of the umbra deep beneath the surface. These magnetic forces are very small: the value of $\beta^{-1}$ is of the order of $10^{-4}$ or less. The thermodynamic parameters at the base of the tube are, roughly speaking, the same as the external parameters scaled by the factor $(1 - \beta^{-1})$ or its powers. Therefore, the internal stratification at $z_b$ has practically the same properties as the stratification of the solar convection zone has at the same depth. Particularly important is the ability of matter to transport energy by convection that is enhanced in the partial ionization zones due to the increased heat capacity of convective elements. This has an effect on the value of the heat flux at the bottom $F_{ub}$ calculated in our model. At first, let us briefly discuss only this property.

The total heat flux is the sum of the radiative and of the convective flux. Within the mixing length theory (*mlt*) it can be written in the following form with a notation basically by Paczyński (1969):

$$F = \frac{16\sigma T^4}{3\kappa \varrho H_p} \nabla + \frac{1}{4\sqrt{2}} c_p \varrho T \alpha^2 \sqrt{gQH_p} (\nabla - \nabla')^{\frac{3}{2}}, \quad (B1)$$

where $\nabla$ is the actual temperature gradient, $\nabla'$ is the gradient of the convective element, $Q = -(\partial \ln \varrho / \partial \ln T)_P$ is the expansion coefficient at constant pressure, and $\kappa$ is the opacity coefficient. Other symbols have their usual meaning. The factors in front of $\nabla$ and $(\nabla - \nabla')^{3/2}$ are the coefficients of the radiative ($\kappa_{\rm rad}$) and of the convective ($\kappa_{\rm con}$) conductivity in the same physical units. These coefficients are functions of the thermodynamic state of the plasma (excluding the parameter $\alpha$ in $\kappa_{\rm con}$), so that they describe the potential ability of the matter to transport heat by the two mechanisms.

Note, that the convective efficiency can be determined only if the heat flux to be transported is known. Then, the *mlt* equations can be solved for the turbulent velocity $v_t$ (which is related to $(\nabla - \nabla')$), and the convective efficiency $\Gamma$, defined as the ratio of the 'heat excess of the element just before dissolving' to the 'heat loss due to radiation', can be related to the ratio $\kappa_{\rm con}/\kappa_{\rm rad}$:

$$\Gamma = 2\sqrt{2} \frac{\kappa_{\rm con}}{\kappa_{\rm rad}} \frac{v_t}{\alpha} (gQH_p)^{-\frac{1}{2}} \quad (B2)$$

(for details see e.g. Cox & Giuli, 1968). The variation of $\Gamma$ and $\kappa_{\rm con}/\kappa_{\rm rad}$ with depth in the solar convection zone model (with $\alpha = 1$) are drawn in the Fig. B1 (see also Fig. 5). That illustrates the partial ionization effects on the heat capacity of the convective element and on the convective efficiency that can be achieved for the specified heat flux value of $F_\odot$.

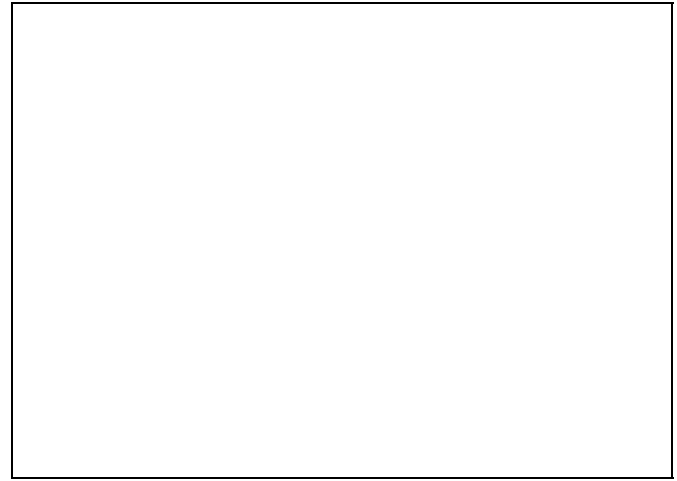

**Fig. B1.** Convective efficiency $\Gamma$ (continuous line), and the ratio of the convective to the radiative conductivity $\kappa_{\rm con}/\kappa_{\rm rad}$ (dashed line) in the solar convection zone

It is also worthwhile to note the negligible role of the radiative diffusion in transporting energy at large depths inside the flux tube (or outside). Even just beneath the surface of the umbra the radiation transports at most 10% of the heat flux. For pure radiative transport the umbra should emit a flux of about $0.02 F_\odot$, which equals only $\sim 10\%$ of the observed irradiation of about $0.2 F_\odot$. At the base of the spot model practically the whole heat is transported by convection and the first term in the Eq. (B1) is negligible.

Now, consider the bottom of the flux tube model, at which the heat flux $F_{ub}$ is not known a priori, but the thermodynamic parameters are implied by the assumed value of $\beta = 8\pi P / B_b^2$.

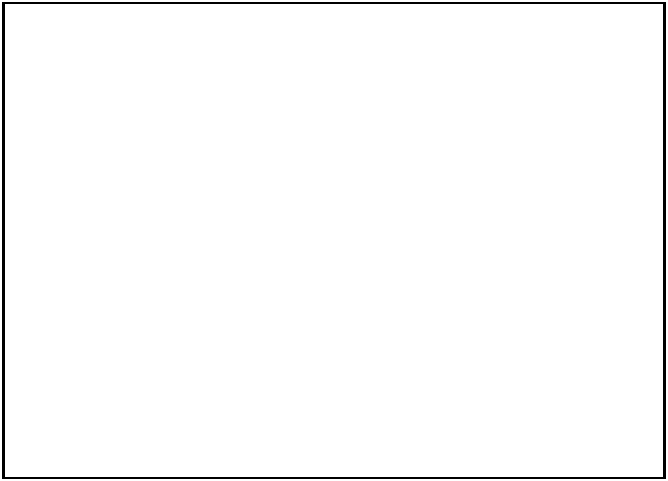

**Fig. B2.** The heat flux $F_{ub}$ that might be transported at the base of the flux tube, if the base was placed at the depth $z_b$, as implied by the lower boundary conditions of the spot model. The curves are parametrized by a constant value of the mixing length parameter. The maximum $F_{ub}$ coincides with a maximum of $\kappa_{con}/\kappa_{rad}$ (cf. Fig. B1). The field strength assumed at the base is constant and equal to 5000 gauss

We can calculate the heat flux that might be transported at each position $z_b$ of the bottom by means of the Eq. (B1) supplemented by the third order equation for the turbulent velocity (see e.g. Paczyński, 1969). The only free parameter we are left with is the mixing length parameter $\alpha$. All other functions are determined by an appropriate scaling of the stratification of the solar convection zone by the value of $\beta$ (i.e. these are functions of $P_e$, $T_e$, and $\beta$).

Figure B2 illustrates the dependence $F_{ub}(z_b)$ calculated in this way for the same value of the field strength $B_b = 5000$ gauss assumed at each depth, and parametrized by several values of $\alpha$. It is natural that $F_{ub}$ is increased at depths where the heat capacity of convective elements is increased, i.e. in the He II ionization zone at $z_b \approx -15$ Mm, and in the zone of H I and He I partial ionization (which is less important for our present models). The local enhancement of $F_{ub}$ may be suppressed when a very small value of $\alpha$ is assumed, because then the convective flux (proportional to $\alpha^2$) is decreased. But this can be done only for a very small total heat flux $F_{ub}$ (smaller than $\sim 0.05 F_\odot$, cf. Fig. B2), since only then the contribution of the radiative flux becomes comparable with the convective flux, so that the enhanced heat capacity becomes less important for the energy transport. In conclusion, we can say that the largest heat flux transported by a purely vertical convection in the flux tube model can be achieved when the base is placed in the middle of the He II ionization zone ($z_b \approx -15$ Mm).

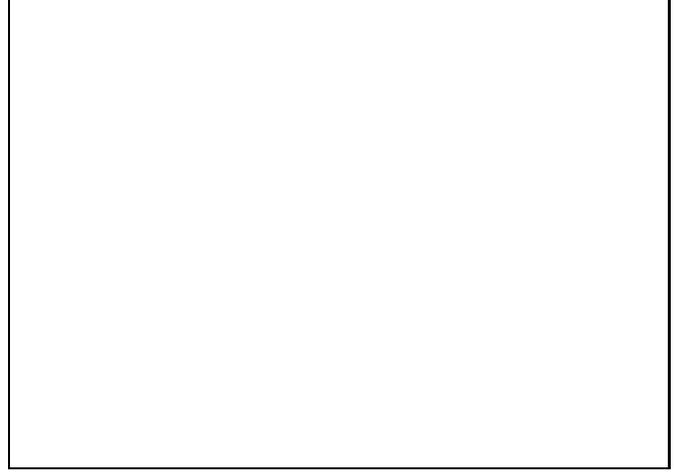

**Fig. B3.** Contours of constant mixing length parameter $\alpha$ in the plane $(F_{ub}, z_b)$, as derived from the lower boundary conditions supplemented by a simple prescription of the channelling of the heat flux (see Eq. (B3)). The behaviour of $F_{ub}(z_b)$ here is very similar to the behaviour of the compressed heat flux in the sunspot models (cf. Fig. 5) for the relevant range of depths $z_b$

Finally, let us take into account the divergence of the heat flux with height, and prescribe the channelling in a simplified way. The umbral luminosity ($\mathcal{F} = \pi R^2 F(z)$), and the umbral magnetic flux (approximated by $\Phi = \pi R^2 B(z)$) are constant with depth, so that we can relate the values of the heat flux and of the field strength at the base to the respective values at the surface:

$$F_{ub}/F_u = B_b/B_c. \qquad (B3)$$

For the surface values we assume that the heat flux $F_u = 0.23 F_\odot$, and the field strength $B_c = 3000$ gauss. Now the field strength $B_b$ at the bottom, or the plasma $\beta$, in the Eq. (B1) is no longer a free parameter, but varies proportionally to the value of the compressed heat flux $F_{ub}$. Again we calculate $F_{ub}$ at each depth $z_b$ by means of the Eq. (B1), in which only $\alpha$ is a free parameter now. The results are plotted in the Fig. B3 which shows contours of a constant parameter $\alpha$. At each depth there are two solutions ($F_{ub}$) for a given value $\alpha$. The lower one (usually $F_{ub} < 0.2 F_\odot$) corresponds to flux tubes that are vertical or only slightly diverge with depth. We do not discuss here neither such solutions nor the layers of $z_b > -10$ Mm, which lie beyond the range of positions $z_b$ in the models. The upper solutions correspond to the channelling discussed in the present spot model. The characteristic shape of $F_{ub}(z_b)$ reflects the mutual dependence of the heat flux convected vertically (as implied by the *mlt*) and of the flux divergence imposed by the Eq. (B3). The relatively largest flux can be transported into the higher layers when the convective ability is maximum ($z_b \approx -15$ Mm, cf. Fig. B2). Then, the divergence of the heat flux required to maintain the umbral luminosity constant with depth may be the smallest. For any displacement of the base position $z_b$, upwards or downwards, the ability of plasma to convect energy decreases, so that the flux convected becomes smaller. Thus, the divergence of the heat flux must be increased, in order to keep the luminosity constant with depth. Therefore, the compression of $F_{ub}$ must be larger. This feature is exhibited in our models (cf. Fig. 5). It is noteworthy, that the magnetostatic equilibrium has a secondary effect on the dependence $F_{ub}(z_b)$. Its pronounced minimum at $z_b = -15$ Mm is caused first of all by the properties of the gravitationally stratified and partially ionized plasma.